\documentclass[twocolumn,twocolappendix]{aastex63}

\usepackage{amsmath}

\received{2021 February 10}
\revised{2021 April 13}
\accepted{2021 April 22}
\published{2021 July 1}
\submitjournal{\aj}

\shorttitle{Perihelion Gap}
\shortauthors{Oldroyd \& Trujillo}

\graphicspath{{./}{figures/}}

\begin{document}

\title{Outer Solar System Perihelion Gap Formation Through Interactions with a\\ Hypothetical Distant Giant Planet}

\correspondingauthor{William J. Oldroyd}
\email{woldroyd@nau.edu}

\author[0000-0001-5750-4953]{William J. Oldroyd}
\affiliation{Department of Astronomy \& Planetary Science\\ Northern Arizona University\\ PO Box 6010, Flagstaff, AZ 86011, USA}

\author[0000-0001-9859-0894]{Chadwick A. Trujillo}
\affiliation{Department of Astronomy \& Planetary Science\\ Northern Arizona University\\ PO Box 6010, Flagstaff, AZ 86011, USA}

\begin{abstract}
Among the outer solar system minor planet orbits there is an observed gap in perihelion between roughly 50 and 65 au at eccentricities $e\gtrsim0.65$. Through a suite of observational simulations, we show that the gap arises from two separate populations, the Extreme Trans-Neptunian Objects (ETNOs; perihelia $q\gtrsim40$ au and semimajor axes $a\gtrsim150$ au) and the Inner Oort Cloud objects (IOCs; $q\gtrsim65$ au and $a\gtrsim250$ au), and is very unlikely to result from a realistic single, continuous distribution of objects. We also explore the connection between the perihelion gap and a hypothetical distant giant planet, often referred to as Planet 9 or Planet X, using dynamical simulations. Some simulations containing Planet X produce the ETNOs, the IOCs, and the perihelion gap from a simple Kuiper-Belt-like initial particle distribution over the age of the solar system. The gap forms as particles scattered to high eccentricity by Neptune are captured into secular resonances with Planet X where they cross the gap and oscillate in perihelion and eccentricity over hundreds of kiloyears. Many of these objects reach a minimum perihelia in their oscillation cycle within the IOC region increasing the mean residence time of the IOC region by a factor of approximately five over the gap region. Our findings imply that, in the presence of a massive external perturber, objects within the perihelion gap will be discovered, but that they will be only $\sim20$\% as numerous as the nearby IOC population ($65$ au $\lesssim q \lesssim 100$ au).
\end{abstract}


\section{Introduction} \label{sec:intro}
The discovery of (90377) Sedna suggested the existence of a population of objects well beyond the known extent of the Kuiper Belt \citep{2004ApJ...617..645B}. As the outer edge of the Kuiper Belt was measured to be at $\sim47$ au \citep{2001ApJ...554L..95T}, a mechanism for scattering Sedna outward from the Kuiper Belt or inward from the Oort cloud was needed to explain the distant location ($q=76.26$ au) of this new object. Some of the hypotheses proposed included single stellar encounters, migration through the stellar birth cluster, or perturbations from a distant undiscovered planet, and all of these scenarios predict the presence of additional objects at similarly high perihelia \citep{2004ApJ...617..645B,2004Natur.432..598K,2004AJ....128.2564M,2006ApJ...643L.135G,2010Sci...329..187L,2010ApJ...720.1691S}.

The idea that an undiscovered planet could be responsible for the high perihelia of Sedna and other similar objects was revisited shortly after the discovery of another distant object, 2012 VP$_{113}$ \citep{2014Natur.507..471T,2016AJ....151...22B}. Along with announcing its discovery, \cite{2014Natur.507..471T} note that a distant planet could constrain the arguments of perihelion $\omega$ of 2012 VP$_{113}$ and other objects with similarly high semimajor axes and perihelia; these objects have been classified as Extreme Trans-Neptunian Objects (ETNOs) for semimajor axes $a\gtrsim150$ au and perihelia $q\gtrsim40$ au, and Inner Oort Cloud objects (IOCs) for $a\gtrsim250$ au and $q\gtrsim65$ au \citep{2020tnss.book...79T}. In addition to highlighting this apparent orbital alignment, \cite{2014Natur.507..471T} noted that there was strong indication for an inner edge to the IOC distribution. The inner Oort cloud has also been shown to have an inner edge near the location of Sedna in Oort cloud formation simulations through interactions within a dense stellar birth cluster \citep{2012Icar..217....1B}.

Since the discovery of 2012 VP$_{113}$, there have been over a hundred studies conducted exploring the possible orbital parameters of the planet \citep{2014Natur.507..471T,2016AJ....151...22B,2017AJ....154...61B,2019AJ....157..139S,2020AJ....159..285C}, its location in the sky \citep{2016ApJ...824L..23B,2016AJ....152...80H,2016ApJ...824L..22M,2016AJ....152..221S,2017AJ....153...91M,2020A&A...640A...6F}, and the implications an undiscovered distant giant planet would have for our solar system \citep{2016AJ....152..126B,2016ApJ...833L...3B,2017AJ....154..229B,2017ApJ...845...27N,2017AJ....154...62V,2018AJ....156..263L,2020ApJ...899L..24S}. This planet is often referred to as Planet 9 or Planet X; for more information, see review articles by \cite{2019PhR...805....1B} and \cite{2020tnss.book...79T} and references therein.

\begin{figure}
  \centering
  \includegraphics[width=\linewidth]{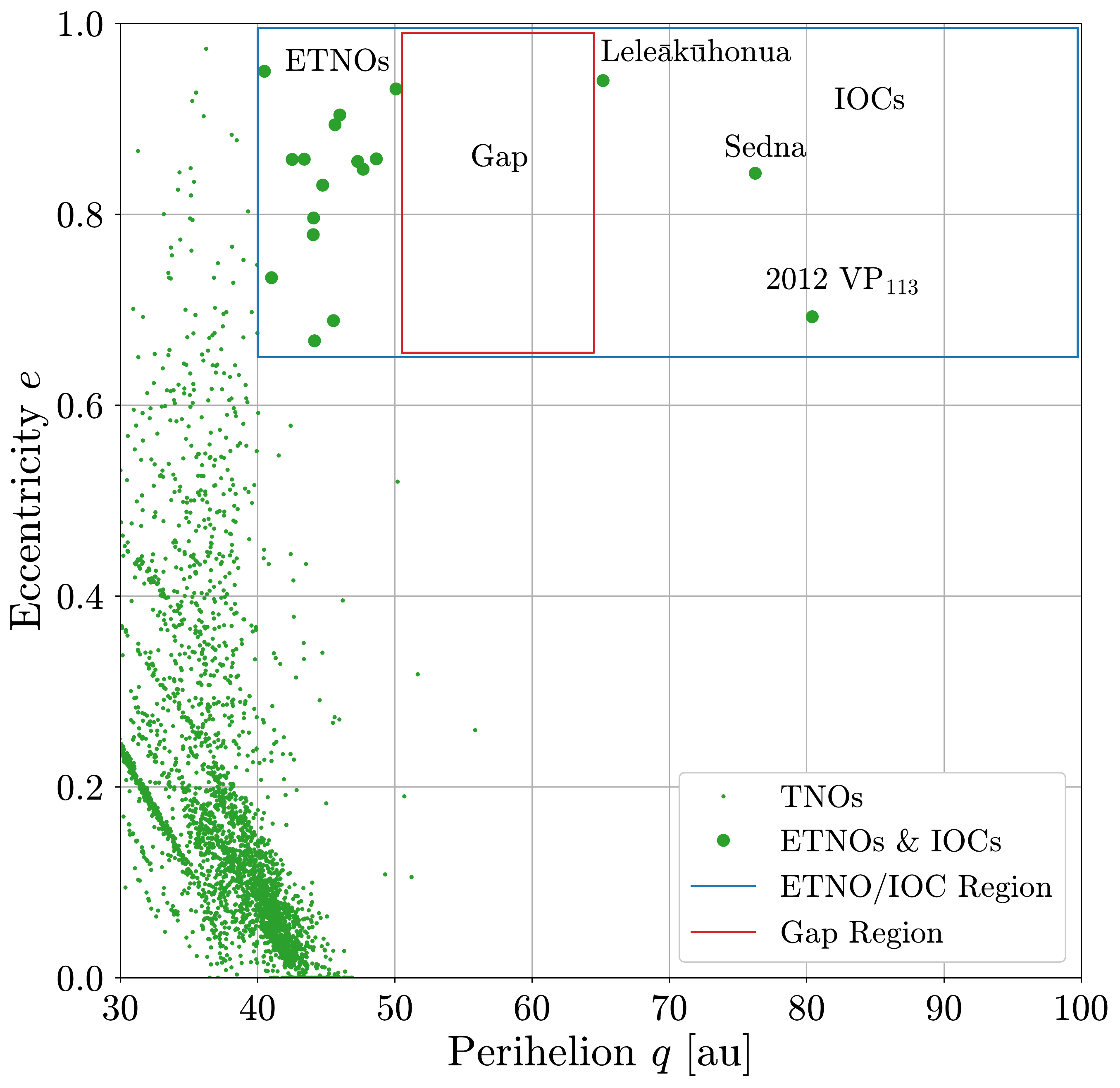}
  \caption{Eccentricity vs. perihelion for all known Trans-Neptunian Objects (TNOs) with perihelion $q>30$ au (green points), circles representing the ETNOs and IOCs are enlarged for emphasis. The perihelion gap (50 au $\lesssim q\lesssim$ 65 au and $0.65\lesssim e<1$) is outlined in red and the ETNO/IOC region (40 au $\lesssim q\lesssim$ 100 au and $0.65\lesssim e<1$) is outlined in blue.} 
  \label{fig:oss}
\end{figure}

Recently, a third IOC has been discovered, (541132) Lele{\={a}}k{\={u}}honua (provisional designation: 2015 TG$_{387}$). Along with this discovery \cite{2019AJ....157..139S} find that the lack of objects discovered with 50 au $\lesssim q \lesssim$ 65 au approaches the $3\sigma$ significance threshold---as objects within this region would be roughly five times easier to detect than the IOCs. They term this the ``gap region" (the red box in Figure \ref{fig:oss}), and state that it may be formed through resonant interactions with Planet X. However, \cite{2020tnss.book...61K} propose that the perihelion gap may instead be a way to rule out the presence of Planet X as simulations that include Planet X scatter objects throughout the gap region rather than preserving this observed structure. Additionally, \cite{2020AJ....160...50Z} show that a feature resembling the perihelion gap can be formed through a self-gravitating disk of primordial planetesimals which undergo an inclination instability if they collectively exceed a mass threshold of 20 Earth masses.

In this work, we examine the agreement between the observed perihelion gap and various distributions for the ETNO/IOC population through observational simulations. We also explore the relationship between the gap and the hypothesised distant giant planet using dynamical simulations and discuss the implications of our results; namely (1) the perihelion gap appears to be the result of a separation between two distinct populations, the ETNOs and the IOCs, (2) some Planet X orbits can form the ETNOs, the IOCs, and the perihelion gap from an initial Kuiper-Belt-like distribution of objects, and (3) this connection between the perihelion gap and Planet X may provide a way to constrain the mass and orbital properties of Planet X.

\section{Observational Simulations} \label{sec:obs}
\subsection{Parameters and Limits} \label{subsec:params}
We examine the statistical agreement between several outer solar system distribution models (power laws and unimodal and bimodal Gaussian distributions in $a$ and $q$) and the observed distribution of ETNOs and IOCs using a suite of observational simulations. In each simulation, we assume a semimajor axis or perihelion distribution for synthetic objects within the region near the gap (the blue box in Figure \ref{fig:oss}; 40 au $\leq q \leq$ 100 au and $0.65\leq e<1$) along with parameter distributions for other orbital and physical characteristics as outlined in Table \ref{tab:obsparams}. The inclination distribution was calculated using the best-fit results from the Deep Ecliptic Survey in the form of a Gaussian, $\mathcal{G}(\mu,\sigma)$, and Lorentzian,  $\mathcal{L}(I,g)$, distribution, both multiplied by the sine of the inclination, $i$:
\begin{equation}
    \begin{split}
        f(i) = &\{(0.75)\mathcal{G}(i,\mu=0,\sigma=2.3)+\\
        &(0.25)\mathcal{L}(i,I=11.2,g=7.0) \}\times \sin{i}, \label{eq:gulbis}
    \end{split}
\end{equation}
where $\mu$ is the mean of the Gaussian, $\sigma$ is the Gaussian standard deviation, $I$ is the Lorentzian half inclination, and $g$ is the Lorentzian power \citep[see][Equation (32) and Table 2]{2010AJ....140..350G}.

The size distribution was calculated as:
\begin{equation}
    r = r_{\text{min}} + (r_{\text{max}} - r_{\text{min}})(1 - \mathbb{R}[0,1]^4),
    \label{eq:rdist}
\end{equation}
where $\mathbb{R}[0,1]$ is the range of real numbers from 0 to 1; the minimum radius $r_{\text{min}}$ was the smallest detectable size for a spherical object in our synthetic survey corresponding to a heliocentric distance $R=40$ au, a geometric albedo $p_{V}=0.1$, and a visual magnitude $V=26$; and the maximum radius $r_{\text{max}}$ was the mean radius of Pluto (see Table \ref{tab:obsparams}). This radius distribution results in few large synthetic objects and many small ones. Using different power laws for the size distribution had negligible effects on the results.

When particles were drawn from a semimajor axis $a$ distribution, perihelion $q$ was calculated as $q=a(1-e)$. When particles were instead drawn from a perihelion distribution, their semimajor axes were calculated using $a=q/(1-e)$. Power-law distributions in $q$ were calculated as:
\begin{equation}
    q = q_{\text{min}} + (q_{\text{max}} - q_{\text{min}})(1 - \mathbb{R}[0,1]^x),
    \label{eq:qdist}
\end{equation}
where $q_{\text{min}}$ and $q_{\text{max}}$ are the perihelion limits defined in Table \ref{tab:obsparams} and $x$ is a real valued exponent ranging from 1 to 4. Power-law distributions in $a$ were calculated using Equation (\ref{eq:qdist}) as well, but with $a$ replacing $q$ in each instance.

\begin{deluxetable}{ccc}
\tablenum{1}
\tablecaption{Observational Simulation Parameters\label{tab:obsparams}}
\tablewidth{0pt}
\tablehead{
\colhead{Parameter} & \colhead{Range} & \colhead{Distribution}
}
\startdata
eccentricity $e$ & $0.65\leq e<1$ & uniform\\
inclination $i$ & $0^{\circ}\leq i\leq90^{\circ}$ & composite$^\text{a}$\\
argument of & $0^{\circ}\leq \omega<360^{\circ}$ & uniform\\
perihelion $\omega$ &&\\
longitude of & $0^{\circ}\leq \Omega<360^{\circ}$ & uniform\\
ascending node $\Omega$ &&\\
mean anomaly $M$ & $0^{\circ}\leq M<360^{\circ}$ & uniform\\
object radius $r$ & $25\leq r\leq1188$ [km] & power-law$^\text{b}$\\
geometric albedo $p_V$ & $p_V=0.1$ & constant\\
perihelion $q$ & $40\leq q\leq100$ [au] & varies$^\text{c}$\\
semimajor axis $a$ & $40\leq a\leq2000$ [au] & varies$^\text{c}$\\
\enddata
\tablecomments{\\$^\textrm{a}$ Determined using Equation (\ref{eq:gulbis}), \citep[see][]{2010AJ....140..350G}.\\ $^\textrm{b}$ The size distribution was calculated using Equation (\ref{eq:rdist}) and limits correspond to the smallest observable object within the simulated survey limits (see Table \ref{tab:obslimits}) and the radius of Pluto.\\ $^\textrm{c}$ When particles were drawn from a distribution in $a$, $q$ was calculated and vice versa.}
\end{deluxetable}

Synthetic objects were ``detected" in the simulation if they were within the observationally derived limits outlined in Table \ref{tab:obslimits}. These constraints were selected to simulate the detection limits of a realistic outer solar system survey, including on-sky velocity $v$, heliocentric distance $R$, ecliptic latitude $b$, and faintest observable magnitude $V$. Once the detectability of these objects was determined, the distribution of detected synthetic objects was compared statistically with the true observed distribution.

\begin{deluxetable}{cc}
\tablenum{2}
\tablecaption{Observational Simulation Detection Limits\label{tab:obslimits}}
\tablewidth{0pt}
\tablehead{
\colhead{Parameter} & \colhead{Limits}
}
\startdata
ecliptic latitude $b$ & $ -20^{\circ} < b < 20^{\circ}$\\
heliocentric distance $R$ & 40 au $< R <$ 100 au\\
apparent sky velocity at & 0.\arcsec04 hr$^{-1}$ $< v <$ 4\arcsec hr$^{-1}$\\
opposition $v$ &\\
visual magnitude $V$ & $V < 26$\\
\enddata
\tablecomments{These parameters were calculated using Equations (\ref{eq:limits_start})-(\ref{eq:limits_end}). The limits are derived from constraints typical of modern outer solar system observational surveys.}
\end{deluxetable}

The ecliptic latitude $b$ for a given synthetic object sampled from the above distributions was calculated as
\begin{equation}
    b = \arcsin(\sin(i)\sin(\omega + f)), \label{eq:limits_start}
\end{equation}
\citep[see][Example (6.5)]{fundamental_astronomy} where $i$ is the inclination, $\omega$ is the argument of perihelion, and $f$ is the true anomaly
\begin{equation}
    f = 2\arctan\left(\sqrt{\frac{1+e}{1-e}}\tan\left(\frac{E}{2}\right)\right),
\end{equation}
\citep[Equation (2.46)]{1999ssd..book.....M} with $E$ being the eccentric anomaly. To determine $E$, we utilized the methods outlined in \cite{kepsolve} with Kepler's equation
\begin{equation}
    E - e\sin(E) = M,
\end{equation}
where $M$ is the mean anomaly \citep[Equation (2.52)]{1999ssd..book.....M}. Synthetic objects with $|b|>20^{\circ}$ were excluded from detections similar to the majority of survey fields observed by \cite{2019AJ....157..139S}.

Limits for apparent on-sky velocity and heliocentric distance overlap, particularly for the lower limit for $v$ which is much slower that objects at the upper limit for $R$. However, the upper limit for $v$ was useful for removing fast-moving synthetic objects within the heliocentric distance range. Heliocentric distance, $R$, was computed using the equation
\begin{equation}
    R = \frac{a(1-e^2)}{1+e\cos(f)}
\end{equation}
\citep[Equation (2.19)]{1999ssd..book.....M}. We placed a minimum bound on heliocentric distance at 40 au and set a maximum value for $R$ at 100 au as nearly all TNOs have been discovered at $R<100$ au.\footnote{With two notable exceptions \citep{2018MPEC....Y...14S,farfarout}.}

The apparent on-sky velocity (in arcseconds per hour) at opposition, $v$, was calculated following the method of \cite{1988AJ.....95.1256L}:
\begin{equation}
    v = 148\frac{1-R^{-0.5}}{R-1},
\end{equation}
where $R>1$. This makes the assumption that synthetic objects are discovered at opposition. We set observational limits such that synthetic objects are detected by our simulated survey only if they exhibit on-sky velocities between 0.\arcsec04 and 4\arcsec per hour to rule out objects that require multiple nights between observations to detect movement or objects moving too quickly to be ETNOs.

A final key limit employed in our model is the limiting magnitude of the survey. The apparent magnitude $V$ for synthetic objects is calculated by first determining the absolute magnitude, $H$, of a given synthetic object which we do by rearranging Equation (7) from \cite{1997Icar..126..450H}
\begin{equation}
    H = 5\log_{10}\left(\frac{1329\hspace{2pt}  \text{km}}{2r\sqrt{p_V}}\right),
\end{equation}
where $r$ is the object radius and $p_V$ is the geometric albedo of the object (see Table \ref{tab:obsparams}). Using the $H$ magnitude, we can transform to visual magnitude
\begin{equation}
    V = H + 5\log_{10}(R(R-1)) + C, \label{eq:limits_end}
\end{equation}
where the constant $C=-2.5\log_{10}(2/3)=0.44$ is the phase angle term at opposition (see \cite{fundamental_astronomy}, Equations (7.43) \& (7.48) and \cite{1907Obs....30...96W}, Equation (2)). We chose a magnitude cutoff of $V<26$ as a faintness limit on our simulated survey \cite[similar to the limiting magnitude obtained by][]{2016ApJ...825L..13S}.

Synthetic objects were considered ``observed" if they were within all observational limits (summarized in Table \ref{tab:obslimits}). Each simulation was run until the number of synthetic objects detected in the simulation was equal to the number of known ETNOs and IOCs within the considered region ($N=18$) which are outlined in Table \ref{tab:etnos}. For most distributions in $a$ and $q$, the vast majority of particles were not observable, thus $10^2$ to $10^5$ synthetic objects were generated for each distribution in order to be able to detect the prescribed number.

\begin{deluxetable}{cccc}
\tablenum{3}
\tablecaption{Real ETNOs and IOCs Used\label{tab:etnos}}
\tablewidth{0pt}
\tablehead{
\colhead{Name/Designation} & \colhead{$q$ (au)} & \colhead{$e$} & \colhead{$a$ (au)}
}
\startdata
(90377) Sedna $^*$ & 76.26 & 0.843 & 484.5\\
(148209) 2000 CR$_{105}$ & 44.08 & 0.796 & 216.1\\
(474640) 2004 VN$_{112}$ & 47.29 & 0.856 & 327.5\\
(496315) 2013 GP$_{136}$ & 41.01 & 0.734 & 153.9\\
(505478) 2013 UT$_{15}$ & 44.05 & 0.779 & 198.9\\
(541132) Lele{\={a}}k{\={u}}honua $^*$ & 65.16 & 0.940 & 1085\\
2010 GB$_{174}$ & 48.65 & 0.858 & 342.3\\
2012 VP$_{113}$ $^*$ & 80.40 & 0.693 & 261.5\\
2013 FT$_{28}$ & 43.39 & 0.858 & 304.8\\
2013 RA$_{109}$ & 45.98 & 0.904 & 479.5\\
2013 SY$_{99}$ & 50.08 & 0.931 & 728\\
2014 SR$_{349}$ & 47.68 & 0.847 & 311.9\\
2014 SS$_{349}$ $\dagger$ & 45.29 & 0.697 & 149.5\\
2014 WB$_{556}$ & 42.65 & 0.852 & 288.1\\
2015 KG$_{163}$ & 40.49 & 0.950 & 805\\
2015 KE$_{172}$ $\dagger$ & 44.14 & 0.667 & 132.7\\
2015 RX$_{245}$ & 45.69 & 0.891 & 420.9\\
2018 VM$_{35}$ & 44.96 & 0.822 & 251.9\\
\enddata
\tablecomments{$^*$: Real ETNOs and IOCs used in this work. $\dagger$: these objects technically do not qualify as ETNOs under the definition of \cite{2020tnss.book...79T}, as they have semimajor axes slightly less than 150 au, however, they both lie within the ETNO/IOC region we define in Figure \ref{fig:oss} (40 au $\leq q \leq 100$ au, $0.65 \leq e<1$), so we include them for the purpose of these analyses. These values were retrieved from the JPL Small-Body Database \citep{1996DPS....28.2504G} on 2020 December 22.}
\end{deluxetable}

\subsection{Statistical Analysis} \label{subsec:stats}
\begin{figure}
  \centering
  \includegraphics[width=\linewidth]{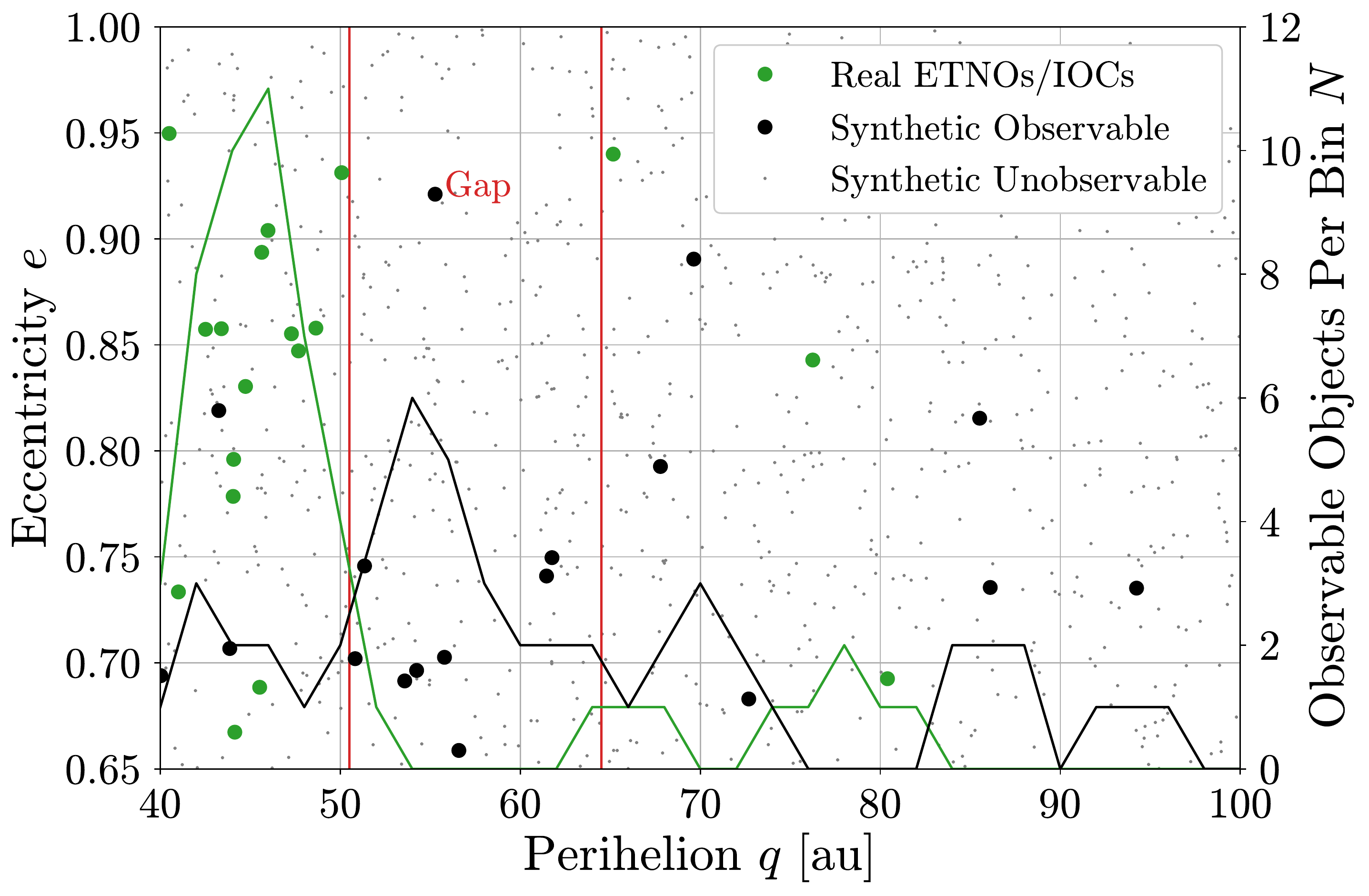}
  \caption{Example rolling histograms of binned perihelion. Real ETNOs and IOCs are shown as green filled circles. Black filled circles indicate synthetic objects that were within the observational limits outlined in Table \ref{tab:obslimits}. Synthetic objects outside these limits are ``unobservable," are shown as gray points, and are not included in further analysis. The perihelion gap is outlined in red. A bin width of 6 au and a bin spacing of 2 au are used. This example is typical of our observational simulations using a single continuous distribution in $a$ or $q$ and synthetic objects appear in the gap region with similar frequency compared to the surrounding regions.}
  \label{fig:histograms}
\end{figure}

To determine the probability that the apparent gap is the result of a random draw from a single continuous distribution, we utilize rolling histograms of perihelion to improve continuity of analysis given the relatively small sample size (see Figure~\ref{fig:histograms}). A variety of bin widths and spacings were tested, all resulting in similar statistical trends. For the analyses presented here, a bin width of 6 au and a bin spacing of 2 au were used, as this combination highlights the features within the gap region. Both real and synthetic objects were binned (separately) and the model was run $10^4$ times for each distribution in semimajor axis and perihelion.

From the results of the model runs, a Poisson maximum likelihood \citep[see][p. 91]{bonamente} was derived for each bin
\begin{equation}
    \hat{\lambda}_i = \frac{1}{n}\sum^n_{j=1}N_{i,j},
\end{equation}
where $n$ is the number of sample runs ($10^4$), and $N_{i,j}$ is the number of particles in the $i$th bin of the $j$th sample run. Using this maximum likelihood and the results from the model runs, we employed two statistical metrics to determine the agreement between the model distributions and the observed distribution: the Poisson Probability, and the Kolmogorov-Smirnov (K-S) Test, as described below.

The Poisson Probability is a measure of how likely the observed distribution is to be drawn from the Poisson maximum likelihood distribution and is computed as
\begin{equation}
    \label{eq:pp}
    P = \exp \sum_{i=1}^{n_{\text{bins}}}\left[\ln{e^{-\hat{\lambda}_i}} + \ln \frac{1}{N_{\text{real},i}!} +
    \ln\hat{\lambda}_i^{N_{\text{real},i}} \right],
\end{equation}
where $n_{\text{bins}}$ is the number of histogram bins, $\hat{\lambda}_i$ is the Poisson maximum likelihood for a given bin, and $N_{\text{real},i}$ is the number of real objects in a given histogram bin \cite[see][p. 91]{bonamente}.

The K-S Test is a metric for determining the likelihood that two data samples are drawn from the same distribution \cite[see][Chapter 14]{press2007numerical}. We implement the \texttt{ks\_2samp} function from the \texttt{scipy.stats python} package \citep{2020SciPy-NMeth} to compare the synthetic objects from all $10^4$ model runs, collectively, with the real objects. The reported metric for these tests is the p value obtained from the analysis.

We ran the model $10^4$ times (sufficient to get a statistical representation; additional runs have minimal effect on the results) for each of 15 distributions (5 in semimajor axis and 10 in perihelion) and calculated the statistical distribution agreement metrics for the results. To compare these single-population distributions with a two-population distribution, we fit the observed distribution with the sum of two Gaussian probability density functions using Markov Chain Monte Carlo (MCMC) statistical sampling described in Section \ref{subsec:mcmc}. The model was then applied to the maximum likelihood fit from this analysis $10^4$ times and the same statistical metrics were calculated for the model results. The results of the statistical analyses are presented in Table \ref{tab:distributions}.

\begin{deluxetable}{cccc}
\tablenum{4}
\tablecaption{Distribution Agreement\label{tab:distributions}}
\tablewidth{0pt}
\tablehead{
\colhead{Parameter} & \colhead{Distribution} & \colhead{$P$} & \colhead{K-S Test}
}
\startdata
$a$ & Uniform & 6.08 e-24 & 0.0006\\
$a$ & Power-law $x=1$ & 1.80 e-27 & $<$0.0001\\
$a$ & Power-law $x=2$ & 4.67 e-30 & $<$0.0001\\
$a$ & Power-law $x=2.7$ & 5.86 e-31 & $<$0.0001\\
$a$ & Power-law $x=4$ & 2.31 e-31 & $<$0.0001\\
$q$ & Uniform & 4.99 e-24 & 0.0005\\
$q$ & Power-law $x=1$ & 2.55 e-20 & 0.0214\\
$q$ & Power-law $x=2$ & 2.14 e-19 & 0.1697\\
$q$ & Power-law $x=3$ & 1.33 e-19 & 0.5484\\
$q$ & Power-law $x=4$ & 9.18 e-21 & 0.7150\\
$q$ & Gaussian $\sigma=5$ & 0 & 0.0053\\
$q$ & Gaussian $\sigma=10$ & 6.10 e-29 & 0.7385\\
$q$ & Gaussian $\sigma=15$ & 3.32 e-21 & 0.3230\\
$q$ & Gaussian $\sigma=20$ & 2.70 e-20 & 0.0735\\
$q$ & Gaussian $\sigma=25$ & 1.13 e-20 & 0.0228\\
\boldmath${q}$ & \textbf{Two-Gaussian fit} & \textbf{6.91 e-12} & \textbf{0.9901}\\
\enddata
\tablecomments{Agreement between the observed ETNO/IOC distribution and modeled synthetic object distributions measured by the Poisson Probability $P$ (Equation (\ref{eq:pp})) and the K-S test. The semimajor axis distribution with a power-law exponent of 2.7 was chosen to match the distribution used in \cite{2019AJ....157..139S} for observational bias calculations. Distributions for other orbital and physical parameters used in these models are given in Table \ref{tab:obsparams}. Single Gaussian distributions all have a mean $\mu=40$ au. The two-Gaussian-fit distribution is the maximum likelihood from the MCMC fit (see Section \ref{subsec:mcmc} and Table \ref{tab:mcmc}). Larger values for the Poisson Probability and K-S test correspond to better distribution agreement between the observed and synthetic objects. Values of 0 indicate that the probability was too low to measure in $10^4$ model runs. Note that the two-Gaussian-fit distribution has significantly better ($\sim 10^7$) agreement with the observed distribution than any other distribution tested.}
\end{deluxetable}

\subsection{Best-fit Distribution} \label{subsec:mcmc}
To compare the effectiveness of a bimodal model, which most resembles the ETNO and IOC populations, with the unimodal distributions described in Section \ref{subsec:params}, we implemented an MCMC statistical sampling fit for the observed object distribution using a two-Gaussian model. To carry out this statistical fit, we utilized the Affine-Invariant Ensemble Sampler from the \texttt{emcee python} package \citep{2013PASP..125..306F}. As a forward model (used for simulating results from the varied parameters at each step that can later be compared with the data), we employed our observational simulation procedure of drawing particles from the distributions given in Table \ref{tab:obsparams}, with a two-Gaussian perihelion distribution (see Table \ref{tab:mcmc}) and the observational limits from Table \ref{tab:obslimits} until the number of synthetic objects detected in the simulation was equal to the number of ETNOs/IOCs in the considered region ($N=18$, see Table \ref{tab:etnos}). This process was repeated 25 times, resulting in averaging over hundreds of particles, for each model step before calculating the Poisson maximum likelihood of the rolling histogram bins corresponding to those used in Section \ref{subsec:stats}.

\begin{deluxetable}{cccc}
\tablenum{5}
\tablecaption{MCMC Parameters and Priors\label{tab:mcmc}}
\tablewidth{0pt}
\tablehead{
\colhead{Parameter} & \colhead{Priors} & \colhead{Model Fit} & \colhead{Max $\mathcal{L}$}
}
\startdata
Gaussian ratio $a_g$ & $0-1$ & $0.59^{+0.10}_{-0.11}$ & 0.62\\
mean 1 $\mu_1$ & $40-50$ au & $44.9^{+0.71}_{-1.25}$ au & 45.0 au\\
standard & $0.1-30$ au & $3.1^{+1.21}_{-0.74}$ au & 2.8 au\\
deviation 1 $\sigma_1$ & & &\\
mean 2 $\mu_2$ & $60-90$ au & $76.9^{+5.83}_{-3.66}$ au & 75.3 au\\
standard & $0.1-30$ au & $9.4^{+9.08}_{-3.40}$ au & 4.9 au\\
deviation 2 $\sigma_2$ & & &
\enddata
\tablecomments{Parameters, priors, and results from the two-Gaussian MCMC fit for the observed ETNO/IOC distribution. The Gaussian ratio is a weight defining the fraction of particles drawn from the first Gaussian versus the second. Priors for Gaussian means were chosen to generously encompass the ETNOs and the IOCs, respectively. The standard deviation priors were also chosen generously to ensure the most likely values would fit comfortably within the range. The model fit is the median of the sample chain distribution with $1 \sigma$ quantiles reported (as is typical for MCMC analyses). The maximum likelihood values $\mathcal{L}$ are the parameters from the single MCMC sample that had the best fit to the data, hence, $\mathcal{L}$ may vary between iterations, but serves as an excellent parameter combination to use for our observational model. A corner plot of the sample distributions reported here is given in Figure \ref{fig:corner}.}
\end{deluxetable}

The natural log of the Poisson Probability (Equation (\ref{eq:pp})) was used at each step as a likelihood function and priors (see Table \ref{tab:mcmc}) were incorporated to give a measurement of the goodness of fit for each model at each step. Short preliminary runs to obtain approximate best-fit values were carried out. The best fit from the short runs was then used as the initialization point for 50 walkers; walkers are statistical sampling chains that pseudo-randomly change their parameters, with weight toward higher likelihoods, at each model step. Each walker was then given a small random offset from the central initialization position to improve the rate at which they explore the parameter space. The model was then run for 15,000 steps which allowed ample time for them to converge (see Appendix \ref{app:mcmc}).

\begin{figure}
  \centering
  \includegraphics[width=\linewidth]{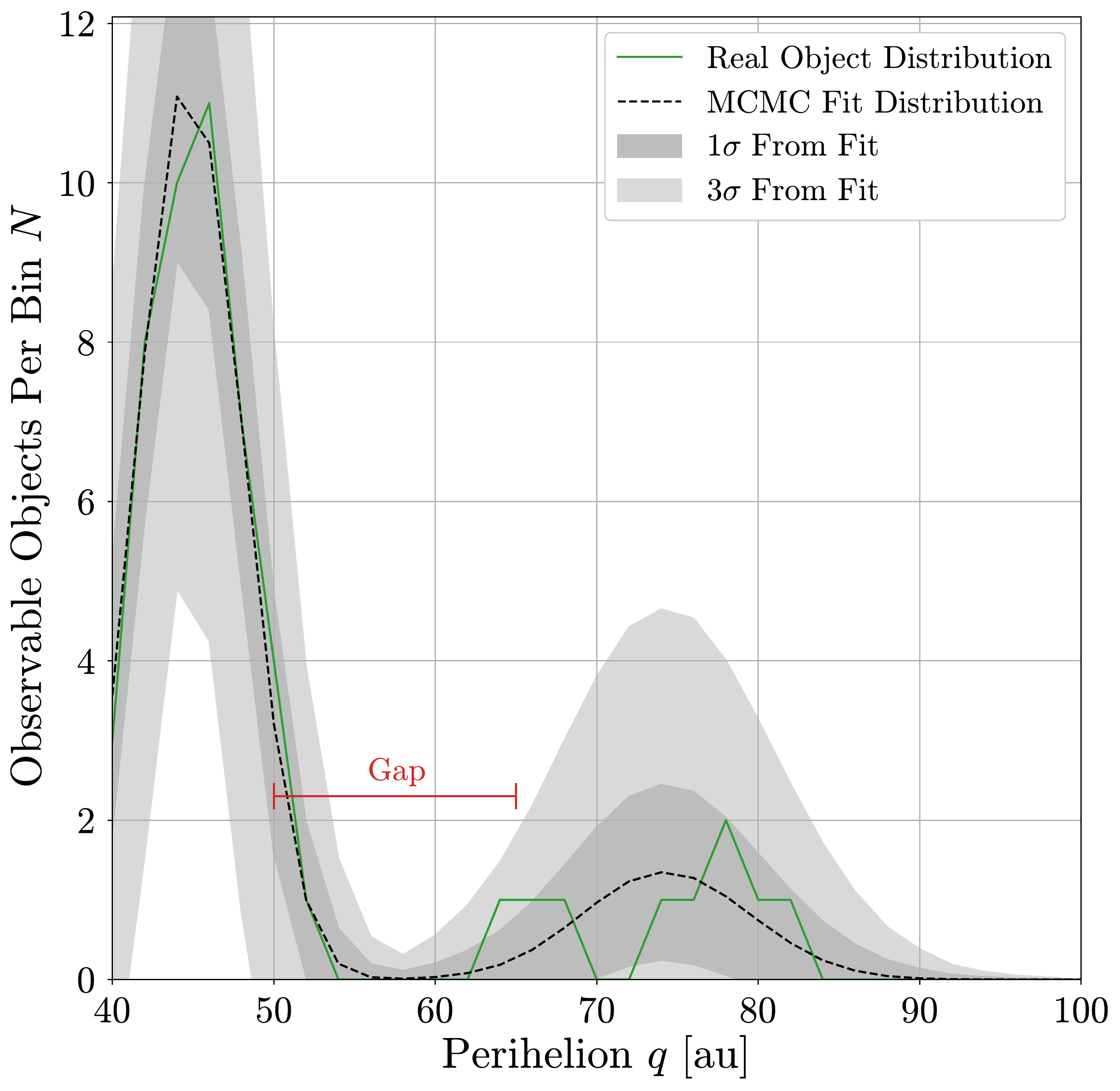}
  \caption{Rolling histogram in perihelion of the observed ETNO/IOC distribution (green) and the two-Gaussian MCMC fit (dashed black line). The dark shaded area indicates the $1 \sigma$ error on the Poisson maximum likelihood and the light shaded area is the $3 \sigma$ error. The extent of the observed perihelion gap is shown for emphasis. Note the excellent agreement between these two distributions.}
  \label{fig:q2gfit}
\end{figure}

Model fit and maximum likelihood parameters from the MCMC sampling are outlined in Table \ref{tab:mcmc}. Maximum likelihood values obtained through the MCMC fit were run through the original observational model $10^4$ times and the statistical tests outlined in Section \ref{subsec:stats} were applied as they were for the other distributions. The two-Gaussian fit resulted in significantly higher distribution agreement than any of the other tested distributions. The results are shown in Table \ref{tab:distributions}, where we see that the two-Gaussian fit is a better fit to the data by $\sim 7$ orders of magnitude. Figure \ref{fig:q2gfit} shows the two-Gaussian fit overlaid on the ETNO/IOC distribution. Here we see a steep reduction in the number of objects at the inner edge of the gap ($\sim$50 au), the perihelion gap with few objects between $\sim$50--65 au, and the IOCs increasing in number beyond $\sim$65 au. 

The improvement in fit of the two-Gaussian fit over the unimodal models indicates that the ETNOs and the IOCs are statistically more than seven orders of magnitude more likely to be two separate populations separated by the gap region rather than the observed distribution developing from a single continuous population. The orbital alignment of the ETNOs and the IOCs has been used as some of the primary evidence for the existence of Planet X \citep{2020tnss.book...79T}. As the separation between these populations defines the perihelion gap, we examine the relationship between Planet X and the ETNOs, IOCs, and ``gap objects" to ascertain whether the Planet X hypothesis is compatible with the presence of the perihelion gap.

\section{Dynamical Simulations} \label{sec:dyn}
To explore the dynamical cause of the separation in $q$ between the ETNOs and IOCs, and the effect an additional distant perturber would have on the perihelion gap, we ran a suite of computational $N$-body simulations. In these simulations, a range of initial test particle distributions interacted with the Sun, the known giant planets, and Planet X.

In each simulation between $10^2$ and $10^4$ test particles were drawn from distributions in $q$ and $e$ to explore the effect of initial particle distribution on the gap region. The test particle distributions in our observational simulations were used for $i$, $\omega$, $\Omega$, and $M$ in our dynamical simulations (see Table \ref{tab:obsparams}). All particles were integrated for 4.5 Gyr with a 0.2 yr time step (roughly 1/60th of the orbital period of Jupiter, the innermost orbit in the simulations), which we found to give an appropriate balance between low amounts of relative energy error (between $\sim10^{-8}$ and $\sim10^{-7}$) and acceptable run times \cite[generally between 24 and 48 hr per simulation, see][]{2015MNRAS.446.1424R}. All integrations were performed using the REBOUND $N$-body integration \texttt{python} package with the hybrid symplectic \texttt{mercurius} integrator designed for both speed and accuracy in complex planetary dynamics scenarios \citep{10.1093/mnras/stz769}.

In order to reduce compute time, dynamical simulations were run in parallel on the Northern Arizona University high-performance computing cluster, \textit{Monsoon}, which has over 2800 cores, peak CPU performance of over 100 teraflops, and is free to use for NAU faculty, staff, and students.\footnote{\url{https://in.nau.edu/hpc/details/}} During integration, particles exceeding a distance threshold ($2\times10^5$ au) were considered ejected and were removed from the simulation to further improve computation time; this is 10 times greater than the limit beyond which \cite{2011Icar..215..491K} noted that $\sim$90\% of objects are ejected over the age of the solar system through galactic tides and stellar perturbations, hence, very few particles beyond this limit would survive and can be ignored.

For each simulation, we divided the results into 10 Myr time steps and computed the average number of particles per region per time step on a grid in perihelion and eccentricity. These results were then plotted as ``heatmaps" (particle density plots) with the $\log_{10}$ of test particle density shown as a color gradient with the real objects overplotted. This facilitates qualitative comparison of the time particles spend in each region between the simulated and observed distributions.

The simulations that were most successful at reproducing the observed features of the ETNOs, the perihelion gap, and the IOCs were initialized with a Kuiper-Belt-like distribution of test particles with 25 au $\leq q \leq$ 40 au and $0\leq e \leq 0.4$. These simulations not only do a better job than other initial distributions we tested at forming a gap-like structure, they also resemble the observed Kuiper Belt. Because our focus is on the region near the gap---the ETNOs, and the IOCs---it is not essential that our simulations exactly reproduce all features of the less extreme TNO distributions (such as the number of objects populating various resonances). Thus, we assume that a simple approximation of a realistic initial TNO distribution, such as the one used here, is sufficient for our investigation.

\begin{figure*}
  \centering
  \includegraphics[width=\textwidth]{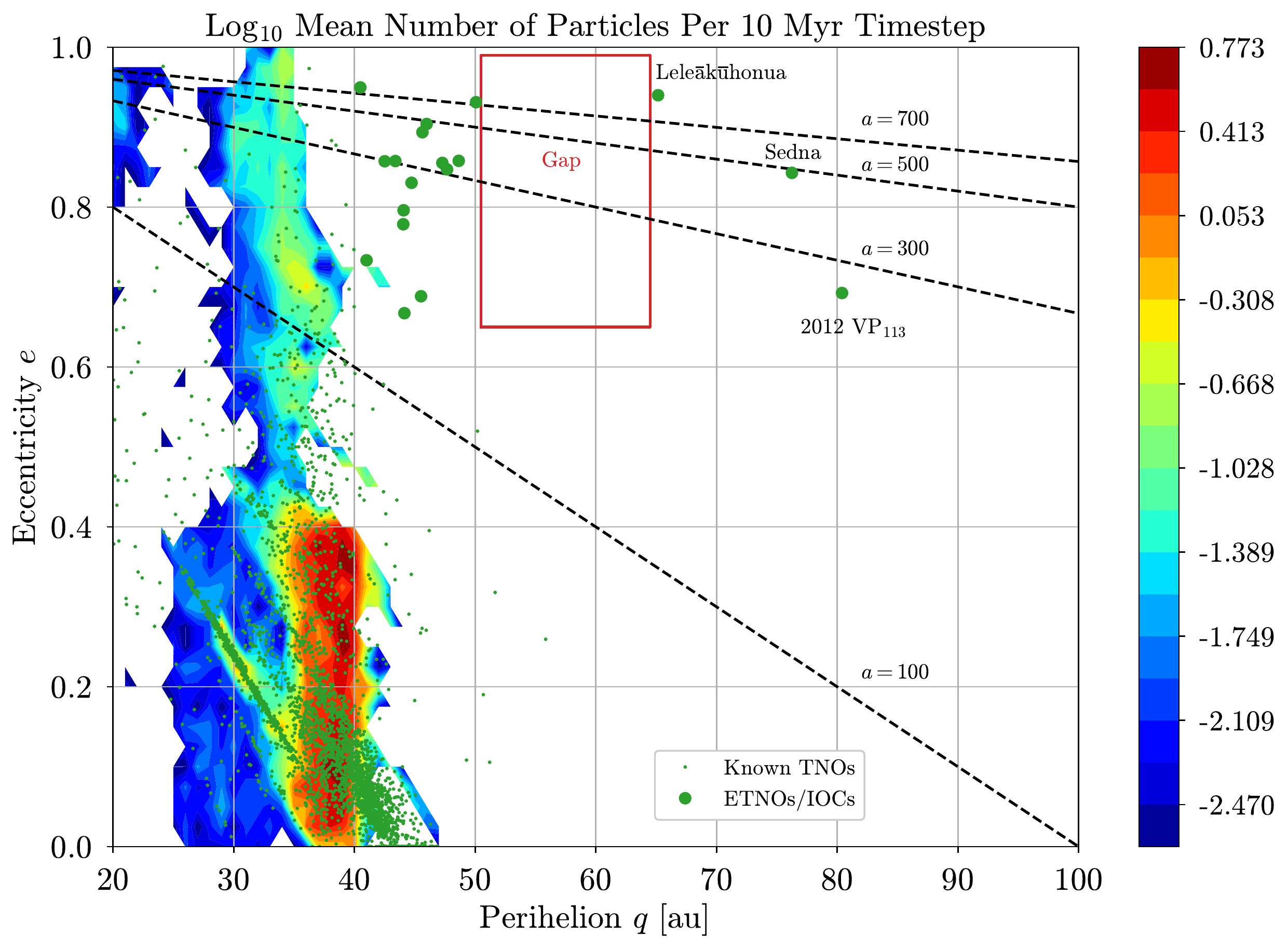}
  \caption{Heatmap in perihelion and eccentricity of the $\log_{10}$ mean number of particles per 10 Myr time step. Known objects are overplotted as green filled circles and ETNOs and IOCs are enlarged for emphasis. Deep reds indicate a constant high particle density and deep blues show that few particles passed through this region over the course of the simulation. White regions had no particle enter the area during the 4.5 Gyr of integration. The gap region and lines of constant $a$ are drawn for reference. This simulation contained the Sun, the giant planets, and an initial Kuiper-Belt-like distribution of $10^3$ test particles with 25 au $\leq q \leq$ 40 au and $0 \leq e \leq 0.4$ and no Planet X. Here we see that the majority of particles remain in a Kuiper-Belt-like region with a small number being scattered to high $e$ by Neptune. No objects, however, are scattered outward to high $q$ near the region of the ETNOs, the perihelion gap, or the IOCs. An animation of the synthetic particles used to create this figure is available in the electronic version of this manuscript (Figure \ref{ani:218}).} 
  \label{fig:hm218}
\end{figure*}

When no Planet X is included in our simulations, a simplified overall Kuiper-Belt-like structure is preserved with the main part of the belt concentrated between roughly 35 au $\lesssim q \lesssim$ 40 au. Several test particles are scattered to high $e$ by Neptune, but there is no scattering to high $q$ in this system, thus, no ETNOs or IOCs are formed and no perihelion gap is present (see Figure \ref{fig:hm218}). These simulation without Planet X, therefore, do not explain the observed features of the distant outer solar system. 

\begin{figure*}
  \centering
  \includegraphics[width=\textwidth]{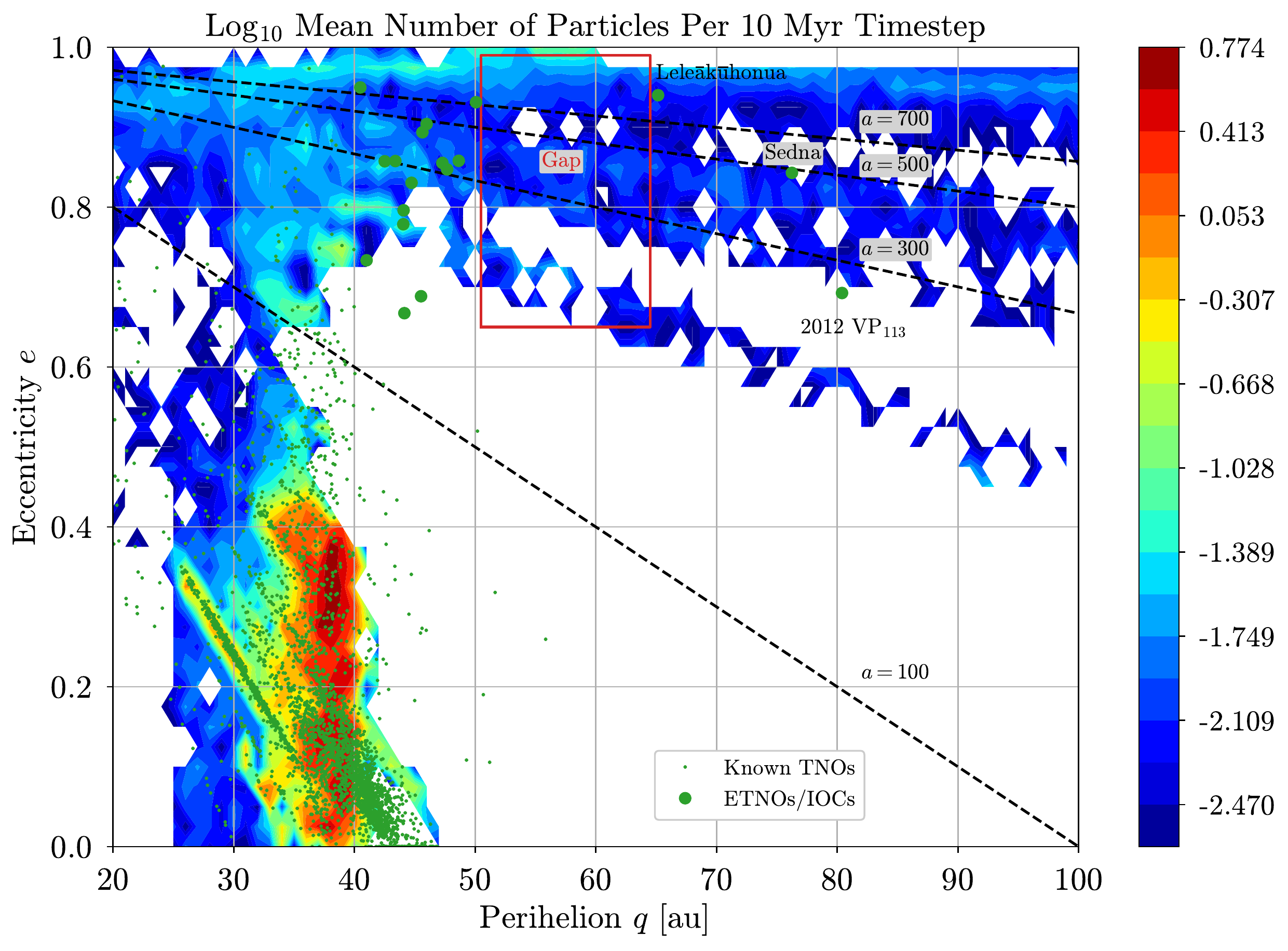}
  \caption{Heatmap in perihelion and eccentricity of the $\log_{10}$ mean number of particles per 10 Myr time step. As Figure \ref{fig:hm218} with the addition of Planet X B19 from Table \ref{tab:pxparams} \citep{2019PhR...805....1B}. Here we see a similar pattern at low $e$ as we observed in the no Planet X case where the majority of particles remain in a Kuiper-Belt-like distribution (see Figure \ref{fig:hm218}). However, with a planet we find that particles are scattered outward to high $q$ by Planet X after they are scattered to high $e$ by Neptune. Additionally, we see a gap-like underdensity, particularly for objects with 0.65 $\lesssim e\lesssim$ 0.8 passing through the perihelion gap. An animation of the synthetic particles used to create this figure is available in the electronic version of this manuscript (Figure \ref{ani:220}).}
  \label{fig:hm220}
\end{figure*}

\begin{figure*}
  \centering
  \includegraphics[width=\textwidth]{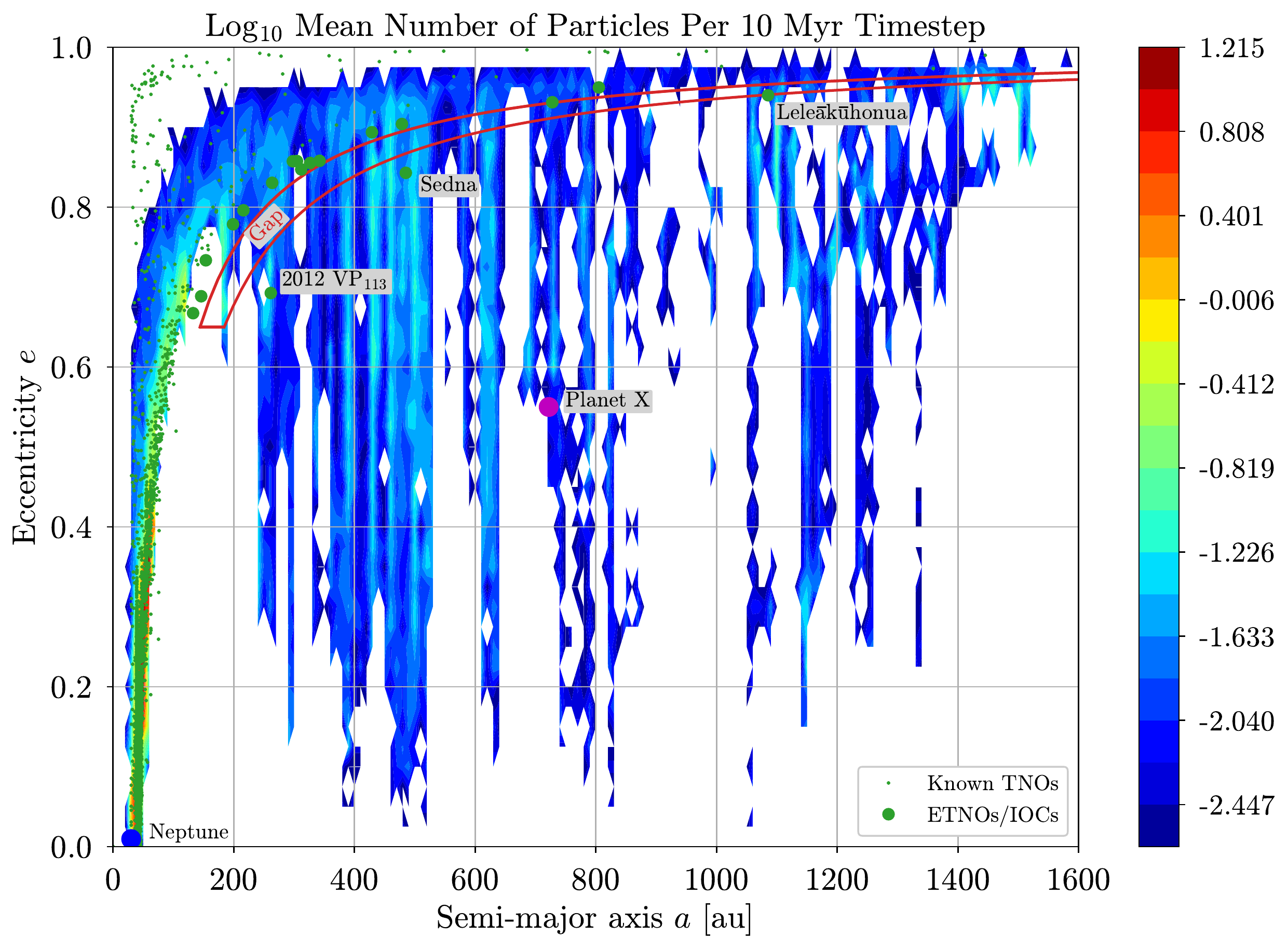}
  \caption{Heatmap in semimajor axis and eccentricity of the $\log_{10}$ mean number of particles per 10 Myr time step. As Figure \ref{fig:hm218} (in $e$ vs. $a$ rather than $e$ vs. $q$) with the addition of Planet X T20 from Table \ref{tab:pxparams} \citep{2020tnss.book...79T}. Neptune and Planet X are shown as large blue and magenta circles respectively. Here we see that objects have their eccentricities dampened as they are scattered to higher perihelia across the gap region along lines of roughly constant semimajor axis by Planet X (the vertical bars). These objects then oscillate between high and low $e$ and many exhibit ``resonance hopping" \citep{2018AJ....156...74B,2020PASP..132l4401K} inducing a slight shift in their semimajor axes. We note that in simulations without Planet X, no objects cross the perihelion gap to low eccentricities (high perihelia). An animation of the synthetic particles used to create this figure is available in the electronic version of this manuscript (Figure \ref{ani:221}).}
  \label{fig:aehm221}
\end{figure*}

Including Planet X in the simulations produced markedly different results. We tested several different candidate parameter sets for Planet X from the literature \citep[e.g.,][see Table \ref{tab:pxparams}]{2019PhR...805....1B,2019EPSC...13.1994B,2020tnss.book...79T}. We find that, with Planet X, an initial test particle distribution uniform in $q$ does not produce a perihelion gap. For Kuiper-Belt-like initial distributions a roughly Kuiper-Belt-like structure is preserved, many Neptune resonances are more readily populated than in the case without Planet X, and particles scattered to high $e$ by Neptune are subsequently scattered to high $q$ by Planet X (see Figure \ref{fig:hm220}). This outward scattering effect happens along lines of roughly constant $a$ as particles are captured into resonance with Planet X. These particles oscillate in eccentricity, in some cases reaching $e<0.1$ (see Figure \ref{fig:aehm221} and the associated animation in Figure \ref{ani:221}). We also note that many of the particles in the ETNO and IOC region in our simulations clearly exhibit ``resonance hopping," as discussed in \cite{2018AJ....156...74B} and \cite{2020PASP..132l4401K}, which accounts for their slight semimajor axis variations seen over secular timescales.

\begin{deluxetable}{cccc}
\tablenum{6}
\tablecaption{Sample Planet X Parameters\label{tab:pxparams}}
\tablewidth{0pt}
\tablehead{
\colhead{Parameter} & \colhead{BB19$^\text{a}$} & \colhead{B19$^\text{b}$} & \colhead{T20$^\text{c}$}
}
\startdata
mass $m$ & 6 $m_{\oplus}$ & 5 $m_{\oplus}$ & 10 $m_{\oplus}$\\
semimajor axis $a$ & 300 au & 500 au & 722 au\\
eccentricity $e$ & 0.15 & 0.25 & 0.55\\
inclination $i$ & $17^{\circ}$ & $20^{\circ}$ & $29^{\circ}$\\
argument of & ... & ... & $142^{\circ}$\\
perihelion $\omega$ &&&\\
longitude of & ... & ... & $93^{\circ}$\\
ascending node $\Omega$ &&&\\
initial true anomaly $f$ & ... & ... & $174^{\circ}$\\
perihelion $q$ & 255 au & 375 au & 325 au\\
\enddata
\tablecomments{These were the three parameter sets for Planet X that we tested most extensively. $m_{\oplus}=$ Earth masses. When values for $\omega$, $\Omega$, and $f$ were not given, we used $0^{\circ}$.\\ $^{\textrm{a}}$ \cite{2019EPSC...13.1994B}.\\ $^{\textrm{b}}$ \cite{2019PhR...805....1B}.\\ $^{\textrm{c}}$ \cite{2020tnss.book...79T}. }
\end{deluxetable}

To explore the effectiveness in which Planet X orbits can produce a perihelion gap, we use the methods outlined in Section \ref{sec:obs}, and compare the number densities of objects for the three Planet X parameter combinations outlined in Table \ref{tab:pxparams} over the ETNO region, the gap region, and the IOC region (40 au $\leq q \leq$ 100 au and $0 \leq e<0.65$; blue box in Figure \ref{fig:oss}). The perihelion distribution of the simulated particles was sampled every 500 Myr and aggregated into an overall distribution for each simulation. We use a similar rolling histogram technique to that outlined in Section \ref{sec:obs}, but here, simulated objects were each given an observability score by assuming a range of sizes with a distribution matching the radius limits in Table \ref{tab:obsparams}. For each simulated object, the observability limits were computed for $10^3$ radii following an $r^4$ distribution between $r_{\text{min}}$ and $r_{\text{max}}$. Each simulated object was then given a weight based on the smallest radius it could have while still satisfying the observational survey limits outlined in Table \ref{tab:obslimits}. These weighted observability scores of the synthetic objects were then binned and scaled to the peak value of the real ETNO and IOC data.

\begin{figure}
  \centering
  \includegraphics[width=\linewidth]{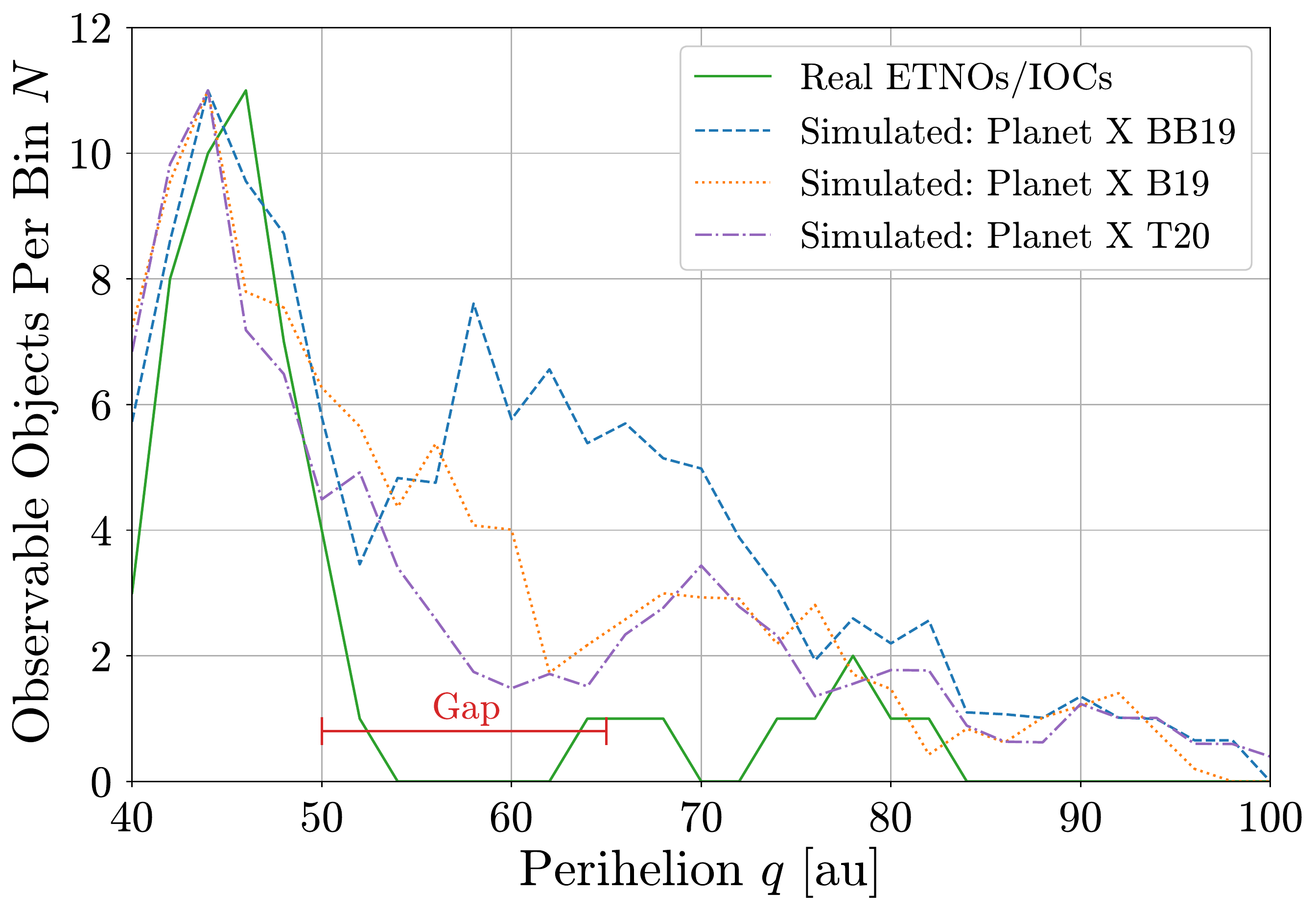}
  \caption{Rolling histogram in $q$ of the real ETNOs and IOCs (solid green) and synthetic objects (dashed blue, dotted orange, and dotted-dashed purple) from dynamical simulations containing Planet X with parameters corresponding to those outlined in Table \ref{tab:pxparams}. Note that the simulated data all exhibit gap-like features. Planet X BB19 causes a deep gap-like feature between $\sim$50 and 56 au, Planet X B19 causes a small dip resembling a gap at roughly 62 au, and Planet X T20 causes a broad gap-like feature from $\sim$56 to 68 au. Additionally, none of these features is the result of an absence of objects within the perihelion gap, but rather is marked by a local minimum in the number of observable objects. We note that we do not favor any particular parameter combination for Planet X given our analyses, but emphasize that many different Planet X orbits may result in the observed perihelion gap. Additionally, simulations that do not include Planet X do not produce ETNOs, Gap Objects, or IOCs and result in zero high eccentricity observable objects for every perihelion value shown in this figure.}
  \label{fig:dynhist}
\end{figure}

The results from this analysis, shown in Figure \ref{fig:dynhist}, indicate that, for some parameter combinations for Planet X (see Table \ref{tab:pxparams}), a feature resembling the perihelion gap is formed. The Planet X BB19 curve has a deep gap-like feature, however, it is offset from and is a poor fit to the observed ETNO/IOC population distribution. Only a weak gap feature is seen in the Planet X B19 curve, but in the Planet X T20 curve, there is a moderate gap feature close to the location of the observed perihelion gap. Hence, there are parameter combinations for Planet X that can form the ETNOs, the IOCs, and features resembling the perihelion gap from simple Kuiper-Belt-like initial distributions. Additionally, we note that we do not favor any particular set of Planet X parameters based on these analyses as we have not evaluated the effect which small differences in parameters may have on the ability of a given Planet X orbit to form a perihelion gap. 

Because of the difference in ability of Planet X orbits to produce the perihelion gap given different parameter combinations, the extent and location of the gap may provide valuable constraints on the mass and orbit of Planet X upon further examination. In particular, the perihelion gap may be used to rule out some, but not all, parameter combinations for Planet X \citep[see][]{2020tnss.book...61K}.

\section{Discussion} \label{sec:discussion}
From the results in Table \ref{tab:distributions}, we find that a bimodal distribution (the two-Gaussian fit) is a better fit, over a single continuous distribution, to the observed ETNOs and IOCs by several orders of magnitude. This is in agreement with previous estimates for the statistical significance of the gap region of nearly 3$\sigma$ \citep{2019AJ....157..139S}. We find all continuous distributions we tested (a wide range that encompasses most realistic possibilities) to be poor fits to the observed distribution of ETNOs and IOCs, and, thus, we find it unlikely that the perihelion gap is the result of a continuous population of distant objects in perihelion or semimajor axis and eccentricity. As more IOCs are discovered, the extent and magnitude of the perihelion gap will become more clear; roughly 10 ETNOs/IOCs discovered in a single well characterized survey (such as the Legacy Survey of Space and Time) would raise the significance above 3$\sigma$ for that survey alone \citep{2020tnss.book...79T}.

By examining our dynamical simulations that do not contain Planet X, we find that neither the ETNO population nor the IOC population can be formed from a simplified Kuiper-Belt-like initial configuration solely by Neptune scattering. An alternative method for forming these populations is Oort cloud diffusion caused by galactic tides and close stellar passages. This mechanism has been shown to have a $\sim$20\%-30\% likelihood of placing objects onto orbits within the ETNO and IOC regions, particularly if the Sun experienced outward radial migration in the galactic disk \citep{2011Icar..215..491K}. However, this method does not appear to produce a perihelion gap, but rather a single population of Oort cloud objects that extends from the ETNO region outward \citep[see Figure 11 of][]{2011Icar..215..491K}. In our dynamical simulations, we focus on the ability of Planet X to cause the perihelion gap through secular interactions and, thus, do not further explore the effects of stellar encounters or galactic tides on the system. Additionally, we assume that interactions with Planet X take place after planet migration and, hence, initial conditions resemble a Kuiper-Belt-like structure.

Initial configurations that instead begin with substantial ETNO and IOC populations separated by a perihelion gap largely retain this structure and preserve the gap over the age of the solar system in the absence of Planet X. This is expected since Neptune only weakly influences objects with $a>60$ au even accounting for mean-motion and Kozai resonances \citep{2008ssbn.book..259G}. Objects in these regions are unlikely to have formed in situ---they would be ejected during planet migration if they formed early on and at later times there was insufficient material to form efficiently at these distances---\citep{1999ApJ...526..465K,2020tnss.book...25M} and we find it is difficult to scatter objects to these regions without the help of a massive external perturber, such as Planet X  \citep[or a similarly massive collection of distant objects,][]{2020AJ....160...50Z}. In our dynamical simulations that began with an initial Kuiper-Belt-like particle distribution and did not contain Planet X, we found that some objects were scattered to high eccentricities by Neptune, but no objects had their perihelia raised to become ETNOs, ``gap objects," or IOCs.

When Planet X is present in these simulations, it is the dominant gravitational perturber for objects with orbits in the ETNO and IOC regions. Interactions with Planet X cause objects scattered to high eccentricities by Neptune to migrate into the ETNO region ($0.65 \lesssim e<1$ and 40 au $\lesssim q \lesssim$ 50 au) by raising their perihelia. As ETNO perihelia increase, the influence of Neptune on their orbits is diminished, though this also depends on the eccentricities of the objects. Once objects reach a perihelion of roughly 50 au, their perihelia are increased much more quickly by Planet X. It is not surprising, then, that the perihelion gap begins at $q\sim 50$ au, since objects crossing this transition boundary appear to be completely detached from the influence of Neptune and tend to be captured into secular and/or mean-motion resonances with Planet X (e.g., the vertical patterns in Figure \ref{fig:aehm221}).

\begin{figure}
  \centering
  \includegraphics[width=\linewidth]{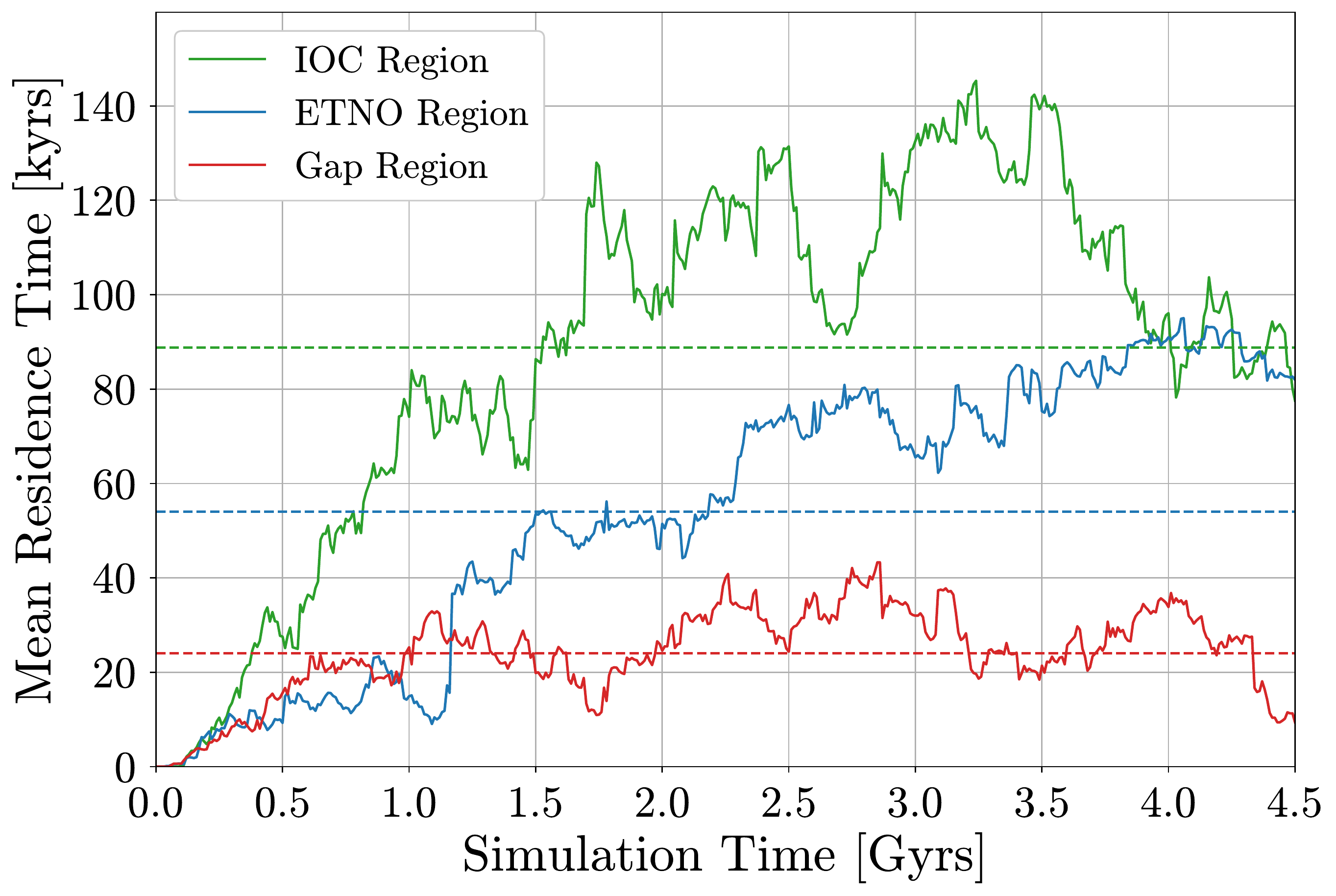}
  \caption{Mean residence time (the time particles spend in a particular region) for the ETNO region (blue), the IOC region (green), and the gap region (red) as a function of simulation time. Horizontal dashed lines show the average residence times of each population over the entire simulation. These data are from a simulation containing Planet X B19 outlined in Table \ref{tab:pxparams}. Note that the average residence time of objects in the IOC region is roughly five times higher than that of particles within the perihelion gap. This difference in mean residence time is the reason for the underdensity of particles within the gap.}
  \label{fig:lifetimes}
\end{figure}

The time it takes objects to cross the perihelion gap depends on the parameters selected for Planet X, with mass being a primary factor. On average, objects spend $\sim$1-2$\times$10$^4$ yr within the gap region in the presence of a 10 Earth mass Planet X and for a 5 Earth mass Planet X they take longer to traverse the gap ($\sim$2-4$\times$10$^4$ yr; see Figure \ref{fig:lifetimes}). After crossing the gap, objects continue to have their perihelia increased by Planet X until they reach their maximum perihelion, which is different for each object and each secular resonance cycle. This increase in perihelion is accompanied by a decrease in eccentricity, hence, the maximum perihelion within a secular cycle for an object corresponds to a minimum eccentricity for that object. These long-term resonant librations in $q$ and $e$ happen along lines of roughly constant semimajor axis (see Figure \ref{fig:aehm221} and the associated animation, Figure \ref{ani:221}).

As these now very distant objects reach their maximum $q$ (minimum $e$) in a libration cycle, their rate of change in $q$ and $e$ slows and then reverses. These objects (``returning IOCs") now decrease in perihelion and increase in eccentricity and migrate back toward the gap. Some of these objects pass back through the IOC region and the perihelion gap and return to the ETNO region. Here, they are again subject to perturbations from Neptune \citep{2020PASP..132l4401K}, and some are ejected from the solar system, but most again reverse direction and begin a new libration cycle as they cross the perihelion gap again and migrate through the IOC region toward a new, and often quite different, maximum $q$ (minimum $e$).

There are several ``returning IOCs," however, that end their libration cycle before crossing back over the gap. These IOCs reach their minimum $q$ within the IOC region (near the location of Sedna and 2012 VP$_{113}$ in $q$-$e$ space) and subsequently begin to increase in $q$ again without reentering the gap region (see Figure \ref{ani:220}, an animation of Figure \ref{fig:hm220}). This process keeps this subset of objects in the IOC region for $\sim10^5$ yr, much longer than the time it takes for objects to traverse the gap. The difference in the time it takes for objects to cross the perihelion gap and how long objects spend at the peak of a libration cycle within the IOC region causes a measurable difference in the number density of objects in the gap region and the IOC region. This difference in mean residence time of objects within the IOC and gap regions is evident in Figure \ref{fig:lifetimes} where particles spend roughly five times longer on average in the IOC region than they do in the perihelion gap.

The mean residence times plotted in Figure \ref{fig:lifetimes} are computed by calculating continuous periods of time that each object spends in each region (e.g., a particle may spend 50 kyr in the ETNO region, followed by 20 kyr in the gap region, then 100 kyr in the IOC region, and subsequently the next 500 kyr beyond the IOC region, then 80 kyr in the IOC region again, and so on). Then, for each time in the simulation, the time particles spend in a particular region is averaged over the number of particles within that region at that simulation time. This metric allows for easy visualization of the difference in the amount of time particles spend in the gap compared to the surrounding regions, which is the reason for the underdensity of particles within the gap.

The slow passage of very high eccentricity objects ($0.9 \lesssim e<1$) through the gap may make it difficult to determine whether or not these objects are temporarily in the gap, or if they are in the IOC region (see high $e$ objects within the gap in Figure \ref{ani:220}, an animation of Figure \ref{fig:hm220}). Combined with the small sample size of known IOCs (Table \ref{tab:etnos}), this slow movement makes determining the outer edge of the perihelion gap quite challenging. Additionally, the ratio of objects in the IOC region to objects within the gap is unknown and depends on the parameters of Planet X and the initial distribution of objects.

As more ETNOs and especially IOCs are discovered through outer solar system surveys, constraints on the perihelion gap and, consequently, Planet X will improve. Our dynamical simulations highlight that the IOCs are indeed the inner edge of a large distant population of objects. Many of these objects are unobservable with current ground and space based telescopes, but advances in observational facilities will continue to expand our capacity to probe the far reaches of our solar system. We predict as these new objects are discovered that some will be discovered within the perihelion gap, however, these ``gap objects" will only be on the order of $\sim$20\% as numerous as the IOCs within 65 au $\lesssim q \lesssim$ 100 au given the majority of probable parameter combinations for Planet X.

\section{Summary} \label{sec:summary}
In this work, we explore the outer solar system perihelion gap (see Figure \ref{fig:oss}) and its connection to the hypothetical Planet X. We find that (1) the gap is very unlikely to result from a realistic single, continuous distribution but is rather a transition region between two separate populations in the outer solar system, the ETNOs and the IOCs; (2) Neptune cannot form the perihelion gap on its own; and (3) Planet X can form the ETNOs, the IOCs, and the perihelion gap from a simple Kuiper-Belt-like initial distribution through a difference in the mean residence time of ``gap objects" and ``returning IOCs." We predict that, in the presence of Planet X, ``gap objects" will be discovered, but that there will be roughly five times more IOCs discovered with 65 au $\lesssim q \lesssim$ 100 au than ``gap objects."

 We began our study by conducting a series of observational simulations. These simulations contain synthetic objects drawn from several different distributions in perihelion $q$ or semimajor axis $a$ within the ETNO/IOC (Extreme Trans-Neptunian Object/Inner Oort Cloud object) region (see Table \ref{tab:obsparams}). Using observational detection limits typical of modern outer solar system observational surveys, we determine which synthetic objects would be ``observable" given these limits (see Table \ref{tab:obslimits} and Figure \ref{fig:histograms}). Next, we use the Poisson Probability and the K-S test to compare the likelihood of observing the perihelion gap given each distribution. Additionally, we fit a two-Gaussian model to the real ETNO/IOC distribution with Markov Chain Monte Carlo statistical fitting techniques (see Figure \ref{fig:q2gfit} and Table \ref{tab:mcmc}). We find that a bimodal distribution, i.e., a separation between the ETNO and IOC populations in the form of a perihelion gap between them, is a much better fit to the observed ETNO/IOC distribution compared to a unimodal distribution by several orders of magnitude (see Table \ref{tab:distributions}). This suggests that the perihelion gap is not the result of observational bias and that the ETNOs and IOCs are two separate but related populations of objects.

With the assurance that the perihelion gap is a real feature of the outer solar system, we examine its relationship with Planet X through a series of dynamical simulations. In these simulations, we draw particles from similar parameter distributions to those used in the observational simulations, but initialized them in a Kuiper-Belt-like disk. These simulations were then integrated for 4.5 Gyr. When Planet X was not present, particles were scattered to high $e$ by Neptune, but never moved into the ETNO/IOC region (see Figure \ref{fig:hm218}). When Planet X was present, particles scattered to high $e$ by Neptune are subsequently captured into resonance with Planet X and transported across the gap to high $q$ along lines of roughly constant semimajor axis (see Figure \ref{fig:hm220}). This process begins near a perihelion of 50 au coinciding with the inner edge of the perihelion gap.

Once objects cross the perihelion gap from the ETNO region to the IOC region and beyond, they begin secular oscillation in eccentricity and perihelion (see Figure \ref{fig:aehm221}). This oscillation is somewhat stochastic as the extremities of this motion are not fixed, but change each cycle. When these objects reach their minimum eccentricity  in their oscillation cycle, their eccentricities begin to increase and their perihelia are reduced bringing them back toward the gap. A subset of these ``returning IOCs," however, do not cross back over the perihelion gap, but rather reach their maximum eccentricity and minimum perihelion within the IOC region resulting in a difference in population density within the gap region and in the IOC region (see Figure \ref{fig:dynhist}). This difference in population density is augmented by the slow reversal of the direction of eccentricity-perihelion oscillation of the ``returning IOCs" which causes them to have a much longer mean residence time within the IOC region than objects transitioning through the gap region (see Figure \ref{fig:lifetimes}).

From these observations we find that Planet X neatly accounts for the outer solar system perihelion gap through secular resonance librations of ``returning IOCs" and their larger mean residence times over objects within the gap region. Additionally, the connection between Planet X and the perihelion gap may serve as a useful constraint on the mass and orbital properties of Planet X, especially as more ETNOs and IOCs are discovered in the next several years.

\acknowledgments
We thank Renu Malhotra (UArizona) for insights regarding dynamical interactions, Ben Proudfoot (BYU) for advice on animation aesthetics, Garett Brown (UToronto) for suggestions on dynamical simulation integration energy errors, Patrick Tribbett (NAU) and Ty Robinson (NAU) for recommendations regarding MCMC fit probabilities, and our anonymous reviewer who provided helpful feedback which contributed to the improvement of this manuscript. This work was funded through NASA Planetary Astronomy grant 80NSSC18K1006. This work was made possible in part through the State of Arizona Technology \& Research Initiative Program. Some of the computational analyses were run on Northern Arizona University’s Monsoon computing cluster, funded by Arizona’s Technology and Research Initiative Fund. This research has made use of NASA’s Astrophysics Data System. This research has made use of data and/or services provided by the International Astronomical Union's Minor Planet Center.

\vspace{5mm}

\software{
\texttt{astropy} \citep{astropy},
\texttt{corner} \citep{2016JOSS....1...24F},
\texttt{emcee} \citep{2013PASP..125..306F},
\texttt{matplotlib} \citep{4160265},
\texttt{numpy} \citep{harris2020array},
\texttt{pandas} \citep{mckinney-proc-scipy-2010},
REBOUND \citep{2012A&A...537A.128R,10.1093/mnras/stz769},
\texttt{scipy} \citep{2020SciPy-NMeth},
\texttt{tqdm} \citep{casper_da_costa_luis_2020_4293724}.
}

\bibliography{references_arxiv}{}

\begin{thebibliography}{}
\expandafter\ifx\csname natexlab\endcsname\relax\def\natexlab#1{#1}\fi
\providecommand{\url}[1]{\href{#1}{#1}}
\providecommand{\dodoi}[1]{doi:~\href{http://doi.org/#1}{\nolinkurl{#1}}}
\providecommand{\doeprint}[1]{\href{http://ascl.net/#1}{\nolinkurl{http://ascl.net/#1}}}
\providecommand{\doarXiv}[1]{\href{https://arxiv.org/abs/#1}{\nolinkurl{https://arxiv.org/abs/#1}}}

\bibitem[{{Bailey} {et~al.}(2016){Bailey}, {Batygin}, \&
  {Brown}}]{2016AJ....152..126B}
{Bailey}, E., {Batygin}, K., \& {Brown}, M.~E. 2016, \aj, 152, 126

\bibitem[{{Bailey} {et~al.}(2018){Bailey}, {Brown}, \&
  {Batygin}}]{2018AJ....156...74B}
{Bailey}, E., {Brown}, M.~E., \& {Batygin}, K. 2018, \aj, 156, 74

\bibitem[{{Batygin} {et~al.}(2019){Batygin}, {Adams}, {Brown}, \&
  {Becker}}]{2019PhR...805....1B}
{Batygin}, K., {Adams}, F.~C., {Brown}, M.~E., \& {Becker}, J.~C. 2019,
  \physrep, 805, 1

\bibitem[{{Batygin} \& {Brown}(2016{\natexlab{a}})}]{2016AJ....151...22B}
{Batygin}, K., \& {Brown}, M.~E. 2016{\natexlab{a}}, \aj, 151, 22

\bibitem[{{Batygin} \& {Brown}(2016{\natexlab{b}})}]{2016ApJ...833L...3B}
---. 2016{\natexlab{b}}, \apjl, 833, L3

\bibitem[{{Batygin} \& {Morbidelli}(2017)}]{2017AJ....154..229B}
{Batygin}, K., \& {Morbidelli}, A. 2017, \aj, 154, 229

\bibitem[{{Becker} {et~al.}(2017){Becker}, {Adams}, {Khain}, {Hamilton}, \&
  {Gerdes}}]{2017AJ....154...61B}
{Becker}, J.~C., {Adams}, F.~C., {Khain}, T., {Hamilton}, S.~J., \& {Gerdes},
  D. 2017, \aj, 154, 61

\bibitem[{{Bonamente}(2017)}]{bonamente}
{Bonamente}, M. 2017, {Statistics and Analysis of Scientific Data}, 2nd edn.
  (New York: Springer)

\bibitem[{{Brasser} {et~al.}(2012){Brasser}, {Duncan}, {Levison}, {Schwamb}, \&
  {Brown}}]{2012Icar..217....1B}
{Brasser}, R., {Duncan}, M.~J., {Levison}, H.~F., {Schwamb}, M.~E., \& {Brown},
  M.~E. 2012, \icarus, 217, 1

\bibitem[{{Brown} \& {Batygin}(2019)}]{2019EPSC...13.1994B}
{Brown}, M., \& {Batygin}, K. 2019, in EPSC-DPS Joint Meeting 2019 (Geneva:
  Europlanet Society), EPSC--DPS2019--1994

\bibitem[{{Brown} \& {Batygin}(2016)}]{2016ApJ...824L..23B}
{Brown}, M.~E., \& {Batygin}, K. 2016, \apjl, 824, L23

\bibitem[{{Brown} {et~al.}(2004){Brown}, {Trujillo}, \&
  {Rabinowitz}}]{2004ApJ...617..645B}
{Brown}, M.~E., {Trujillo}, C., \& {Rabinowitz}, D. 2004, \apj, 617, 645

\bibitem[{{Clement} \& {Kaib}(2020)}]{2020AJ....159..285C}
{Clement}, M.~S., \& {Kaib}, N.~A. 2020, \aj, 159, 285

\bibitem[{da~Costa-Luis {et~al.}(2020)da~Costa-Luis, Larroque, Altendorf, Mary,
  Korobov, Yorav-Raphael, Ivanov, Bargull, Rodrigues, CHEN, Newey, James,
  Zugnoni, Pagel, mjstevens777, Dektyarev, Rothberg, Alexander, Panteleit,
  Dill, FichteFoll, HeoHeo, van Kemenade, McCracken, Nordlund, Nechaev, Desh,
  RedBug312, richardsheridan, \& Socialery}]{casper_da_costa_luis_2020_4293724}
da~Costa-Luis, C., Larroque, S.~K., Altendorf, K., {et~al.} 2020, {tqdm: A
  fast, Extensible Progress Bar for Python and CLI}, v4.54.0,  Zenodo,
  \dodoi{10.5281/zenodo.4293724}

\bibitem[{{Fienga} {et~al.}(2020){Fienga}, {Di Ruscio}, {Bernus}, {Deram},
  {Durante}, {Laskar}, \& {Iess}}]{2020A&A...640A...6F}
{Fienga}, A., {Di Ruscio}, A., {Bernus}, L., {et~al.} 2020, \aap, 640, A6

\bibitem[{{Foreman-Mackey}(2016)}]{2016JOSS....1...24F}
{Foreman-Mackey}, D. 2016, JOSS, 1, 24

\bibitem[{{Foreman-Mackey} {et~al.}(2013){Foreman-Mackey}, {Hogg}, {Lang}, \&
  {Goodman}}]{2013PASP..125..306F}
{Foreman-Mackey}, D., {Hogg}, D.~W., {Lang}, D., \& {Goodman}, J. 2013, \pasp,
  125, 306

\bibitem[{{Giorgini} {et~al.}(1996){Giorgini}, {Yeomans}, {Chamberlin},
  {Chodas}, {Jacobson}, {Keesey}, {Lieske}, {Ostro}, {Standish}, \&
  {Wimberly}}]{1996DPS....28.2504G}
{Giorgini}, J.~D., {Yeomans}, D.~K., {Chamberlin}, A.~B., {et~al.} 1996, BAAS,
  28, 1158

\bibitem[{{Gladman} \& {Chan}(2006)}]{2006ApJ...643L.135G}
{Gladman}, B., \& {Chan}, C. 2006, \apjl, 643, L135

\bibitem[{{Gomes} {et~al.}(2008){Gomes}, {Fern{\'a}ndez}, {Gallardo}, \&
  {Brunini}}]{2008ssbn.book..259G}
{Gomes}, R.~S., {Fern{\'a}ndez}, J.~A., {Gallardo}, T., \& {Brunini}, A. 2008,
  {in The Solar System Beyond Neptune}, ed. M.~A. {Barucci}, H.~{Boehnhardt},
  D.~P. {Cruikshank}, A.~{Morbidelli}, \& R.~{Dotson} (Tuscon, AZ: Univ.
  Arizona Press), 259

\bibitem[{{Gulbis} {et~al.}(2010){Gulbis}, {Elliot}, {Adams}, {Benecchi},
  {Buie}, {Trilling}, \& {Wasserman}}]{2010AJ....140..350G}
{Gulbis}, A.~A.~S., {Elliot}, J.~L., {Adams}, E.~R., {et~al.} 2010, \aj, 140,
  350

\bibitem[{{Harris} \& {Harris}(1997)}]{1997Icar..126..450H}
{Harris}, A.~W., \& {Harris}, A.~W. 1997, \icarus, 126, 450

\bibitem[{Harris {et~al.}(2020)Harris, Millman, van~der Walt, Gommers,
  Virtanen, Cournapeau, Wieser, Taylor, Berg, Smith, Kern, Picus, Hoyer, van
  Kerkwijk, Brett, Haldane, del R{'{\i}}o, Wiebe, Peterson,
  G{'{e}}rard-Marchant, Sheppard, Reddy, Weckesser, Abbasi, Gohlke, \&
  Oliphant}]{harris2020array}
Harris, C.~R., Millman, K.~J., van~der Walt, S.~J., {et~al.} 2020, Nature, 585,
  357

\bibitem[{{Holman} \& {Payne}(2016)}]{2016AJ....152...80H}
{Holman}, M.~J., \& {Payne}, M.~J. 2016, \aj, 152, 80

\bibitem[{{Hunter}(2007)}]{4160265}
{Hunter}, J.~D. 2007, CSE, 9, 90

\bibitem[{{Kaib} {et~al.}(2011){Kaib}, {Ro{\v{s}}kar}, \&
  {Quinn}}]{2011Icar..215..491K}
{Kaib}, N.~A., {Ro{\v{s}}kar}, R., \& {Quinn}, T. 2011, \icarus, 215, 491

\bibitem[{{Karttunen} {et~al.}(2007){Karttunen}, {Kr\"{o}ger}, {Oja},
  {Poutanen}, \& {Donner}}]{fundamental_astronomy}
{Karttunen}, H., {Kr\"{o}ger}, P., {Oja}, H., {Poutanen}, M., \& {Donner},
  K.~J., eds. 2007, {Fundamental Astronomy}, 5th edn. (Berlin: Springer)

\bibitem[{{Kavelaars} {et~al.}(2020){Kavelaars}, {Lawler}, {Bannister}, \&
  {Shankman}}]{2020tnss.book...61K}
{Kavelaars}, J.~J., {Lawler}, S.~M., {Bannister}, M.~T., \& {Shankman}, C.
  2020, {in The Trans-Neptunian Solar System}, ed. D.~{Prialnik}, M.~A.
  {Barucci}, \& L.~{Young} (Amsterdam: Elsevier), 61

\bibitem[{{Kenyon} \& {Bromley}(2004)}]{2004Natur.432..598K}
{Kenyon}, S.~J., \& {Bromley}, B.~C. 2004, \nat, 432, 598

\bibitem[{{Kenyon} \& {Luu}(1999)}]{1999ApJ...526..465K}
{Kenyon}, S.~J., \& {Luu}, J.~X. 1999, \apj, 526, 465

\bibitem[{{Khain} {et~al.}(2020){Khain}, {Becker}, \&
  {Adams}}]{2020PASP..132l4401K}
{Khain}, T., {Becker}, J.~C., \& {Adams}, F.~C. 2020, \pasp, 132, 124401

\bibitem[{{Kraft} {et~al.}(1991){Kraft}, {Burrows}, \&
  {Nousek}}]{1991ApJ...374..344K}
{Kraft}, R.~P., {Burrows}, D.~N., \& {Nousek}, J.~A. 1991, \apj, 374, 344

\bibitem[{{Levison} {et~al.}(2010){Levison}, {Duncan}, {Brasser}, \&
  {Kaufmann}}]{2010Sci...329..187L}
{Levison}, H.~F., {Duncan}, M.~J., {Brasser}, R., \& {Kaufmann}, D.~E. 2010,
  Science, 329, 187

\bibitem[{{Li} {et~al.}(2018){Li}, {Hadden}, {Payne}, \&
  {Holman}}]{2018AJ....156..263L}
{Li}, G., {Hadden}, S., {Payne}, M., \& {Holman}, M.~J. 2018, \aj, 156, 263

\bibitem[{{Luu} \& {Jewitt}(1988)}]{1988AJ.....95.1256L}
{Luu}, J.~X., \& {Jewitt}, D. 1988, \aj, 95, 1256

\bibitem[{{Malhotra} {et~al.}(2016){Malhotra}, {Volk}, \&
  {Wang}}]{2016ApJ...824L..22M}
{Malhotra}, R., {Volk}, K., \& {Wang}, X. 2016, \apjl, 824, L22

\bibitem[{{M}c{K}inney(2010)}]{mckinney-proc-scipy-2010}
{M}c{K}inney, W. 2010, in Proc. 9th Python in Science Conf., ed. S.~van~der
  {W}alt \& J.~{M}illman (Austin, TX: SciPy), 56

\bibitem[{{Millholland} \& {Laughlin}(2017)}]{2017AJ....153...91M}
{Millholland}, S., \& {Laughlin}, G. 2017, \aj, 153, 91

\bibitem[{{Morbidelli} \& {Levison}(2004)}]{2004AJ....128.2564M}
{Morbidelli}, A., \& {Levison}, H.~F. 2004, \aj, 128, 2564

\bibitem[{{Morbidelli} \& {Nesvorn{\'y}}(2020)}]{2020tnss.book...25M}
{Morbidelli}, A., \& {Nesvorn{\'y}}, D. 2020, {in The Trans-Neptunian Solar
  System}, ed. D.~{Prialnik}, M.~A. {Barucci}, \& L.~{Young} (Amsterdam:
  Elsevier), 25

\bibitem[{Murison(2006)}]{kepsolve}
Murison, M. 2006, A Practical Method for Solving the Kepler Equation, Technical
  Report

\bibitem[{{Murray} \& {Dermott}(1999)}]{1999ssd..book.....M}
{Murray}, C.~D., \& {Dermott}, S.~F. 1999, {Solar System Dynamics} (Cambridge:
  Cambridge Univ. Press)

\bibitem[{{Nesvorn{\'y}} {et~al.}(2017){Nesvorn{\'y}}, {Vokrouhlick{\'y}},
  {Dones}, {Levison}, {Kaib}, \& {Morbidelli}}]{2017ApJ...845...27N}
{Nesvorn{\'y}}, D., {Vokrouhlick{\'y}}, D., {Dones}, L., {et~al.} 2017, \apj,
  845, 27

\bibitem[{Press {et~al.}(2007)Press, Teukolsky, Vetterling, \&
  Flannery}]{press2007numerical}
Press, W., Teukolsky, S., Vetterling, W., \& Flannery, B. 2007, Numerical
  Recipes 3rd Edition: The Art of Scientific Computing (Cambridge: Cambridge
  Univ. Press)

\bibitem[{{Price-Whelan} {et~al.}(2018){Price-Whelan}, {Sip{\H{o}}cz},
  {G{\"u}nther}, {Lim}, {Crawford}, {Conseil}, {Shupe}, {Craig}, {Dencheva},
  {Ginsburg}, {VanderPlas}, {Bradley}, {P{\'e}rez-Su{\'a}rez}, {de Val-Borro},
  {Paper Contributors}, {Aldcroft}, {Cruz}, {Robitaille}, {Tollerud},
  {Coordination Committee}, {Ardelean}, {Babej}, {Bach}, {Bachetti}, {Bakanov},
  {Bamford}, {Barentsen}, {Barmby}, {Baumbach}, {Berry}, {Biscani}, {Boquien},
  {Bostroem}, {Bouma}, {Brammer}, {Bray}, {Breytenbach}, {Buddelmeijer},
  {Burke}, {Calderone}, {Cano Rodr{\'\i}guez}, {Cara}, {Cardoso}, {Cheedella},
  {Copin}, {Corrales}, {Crichton}, {D{\textquoteright}Avella}, {Deil},
  {Depagne}, {Dietrich}, {Donath}, {Droettboom}, {Earl}, {Erben}, {Fabbro},
  {Ferreira}, {Finethy}, {Fox}, {Garrison}, {Gibbons}, {Goldstein}, {Gommers},
  {Greco}, {Greenfield}, {Groener}, {Grollier}, {Hagen}, {Hirst}, {Homeier},
  {Horton}, {Hosseinzadeh}, {Hu}, {Hunkeler}, {Ivezi{\'c}}, {Jain}, {Jenness},
  {Kanarek}, {Kendrew}, {Kern}, {Kerzendorf}, {Khvalko}, {King}, {Kirkby},
  {Kulkarni}, {Kumar}, {Lee}, {Lenz}, {Littlefair}, {Ma}, {Macleod},
  {Mastropietro}, {McCully}, {Montagnac}, {Morris}, {Mueller}, {Mumford},
  {Muna}, {Murphy}, {Nelson}, {Nguyen}, {Ninan}, {N{\"o}the}, {Ogaz}, {Oh},
  {Parejko}, {Parley}, {Pascual}, {Patil}, {Patil}, {Plunkett}, {Prochaska},
  {Rastogi}, {Reddy Janga}, {Sabater}, {Sakurikar}, {Seifert}, {Sherbert},
  {Sherwood-Taylor}, {Shih}, {Sick}, {Silbiger}, {Singanamalla}, {Singer},
  {Sladen}, {Sooley}, {Sornarajah}, {Streicher}, {Teuben}, {Thomas},
  {Tremblay}, {Turner}, {Terr{\'o}n}, {van Kerkwijk}, {de la Vega}, {Watkins},
  {Weaver}, {Whitmore}, {Woillez}, {Zabalza}, \& {Contributors}}]{astropy}
{Price-Whelan}, A.~M., {Sip{\H{o}}cz}, B.~M., {G{\"u}nther}, H.~M., {et~al.}
  2018, \aj, 156, 123

\bibitem[{{Rein} \& {Liu}(2012)}]{2012A&A...537A.128R}
{Rein}, H., \& {Liu}, S.~F. 2012, \aap, 537, A128

\bibitem[{{Rein} \& {Spiegel}(2015)}]{2015MNRAS.446.1424R}
{Rein}, H., \& {Spiegel}, D.~S. 2015, \mnras, 446, 1424

\bibitem[{Rein {et~al.}(2019)Rein, Hernandez, Tamayo, Brown, Eckels, Holmes,
  Lau, Leblanc, \& Silburt}]{10.1093/mnras/stz769}
Rein, H., Hernandez, D.~M., Tamayo, D., {et~al.} 2019, MNRAS, 485, 5490

\bibitem[{{Schwamb} {et~al.}(2010){Schwamb}, {Brown}, {Rabinowitz}, \&
  {Ragozzine}}]{2010ApJ...720.1691S}
{Schwamb}, M.~E., {Brown}, M.~E., {Rabinowitz}, D.~L., \& {Ragozzine}, D. 2010,
  \apj, 720, 1691

\bibitem[{{Sheppard} {et~al.}(2021){Sheppard}, {Tholen}, \&
  {Trujillo}}]{farfarout}
{Sheppard}, S.~S., {Tholen}, D.~J., \& {Trujillo}, C.~A. 2021, Minor Planet
  Electronic Circulars, 2021-C187

\bibitem[{{Sheppard} \& {Trujillo}(2016)}]{2016AJ....152..221S}
{Sheppard}, S.~S., \& {Trujillo}, C. 2016, \aj, 152, 221

\bibitem[{{Sheppard} {et~al.}(2016){Sheppard}, {Trujillo}, \&
  {Tholen}}]{2016ApJ...825L..13S}
{Sheppard}, S.~S., {Trujillo}, C., \& {Tholen}, D.~J. 2016, \apjl, 825, L13

\bibitem[{{Sheppard} {et~al.}(2018){Sheppard}, {Trujillo}, {Oldroyd}, {Tholen},
  \& {Williams}}]{2018MPEC....Y...14S}
{Sheppard}, S.~S., {Trujillo}, C.~A., {Oldroyd}, W.~J., {Tholen}, D.~J., \&
  {Williams}, G.~V. 2018, Minor Planet Electronic Circulars, 2018-Y14

\bibitem[{{Sheppard} {et~al.}(2019){Sheppard}, {Trujillo}, {Tholen}, \&
  {Kaib}}]{2019AJ....157..139S}
{Sheppard}, S.~S., {Trujillo}, C.~A., {Tholen}, D.~J., \& {Kaib}, N. 2019, \aj,
  157, 139

\bibitem[{{Siraj} \& {Loeb}(2020)}]{2020ApJ...899L..24S}
{Siraj}, A., \& {Loeb}, A. 2020, \apjl, 899, L24

\bibitem[{{Trujillo}(2020)}]{2020tnss.book...79T}
{Trujillo}, C. 2020, {in The Trans-Neptunian Solar System}, ed. D.~{Prialnik},
  M.~A. {Barucci}, \& L.~{Young} (Amsterdam: Elsevier), 79

\bibitem[{{Trujillo} \& {Brown}(2001)}]{2001ApJ...554L..95T}
{Trujillo}, C.~A., \& {Brown}, M.~E. 2001, \apjl, 554, L95

\bibitem[{{Trujillo} \& {Sheppard}(2014)}]{2014Natur.507..471T}
{Trujillo}, C.~A., \& {Sheppard}, S.~S. 2014, \nat, 507, 471

\bibitem[{{Virtanen} {et~al.}(2020){Virtanen}, {Gommers}, {Oliphant},
  {Haberland}, {Reddy}, {Cournapeau}, {Burovski}, {Peterson}, {Weckesser},
  {Bright}, {van der Walt}, {Brett}, {Wilson}, {Jarrod Millman}, {Mayorov},
  {Nelson}, {Jones}, {Kern}, {Larson}, {Carey}, {Polat}, {Feng}, {Moore}, {Vand
  erPlas}, {Laxalde}, {Perktold}, {Cimrman}, {Henriksen}, {Quintero}, {Harris},
  {Archibald}, {Ribeiro}, {Pedregosa}, {van Mulbregt}, \&
  {Contributors}}]{2020SciPy-NMeth}
{Virtanen}, P., {Gommers}, R., {Oliphant}, T.~E., {et~al.} 2020, Nature
  Methods, 17, 261

\bibitem[{{Volk} \& {Malhotra}(2017)}]{2017AJ....154...62V}
{Volk}, K., \& {Malhotra}, R. 2017, \aj, 154, 62

\bibitem[{{Whitmell}(1907)}]{1907Obs....30...96W}
{Whitmell}, C.~T. 1907, Obs., 30, 96

\bibitem[{{Zderic} \& {Madigan}(2020)}]{2020AJ....160...50Z}
{Zderic}, A., \& {Madigan}, A.-M. 2020, \aj, 160, 50

\end{thebibliography}
\bibliographystyle{aasjournal}

\restartappendixnumbering
\appendix
\section{MCMC Convergence and Distributions} \label{app:mcmc}

To measure the convergence of our MCMC sampling run of the two-Gaussian distribution, we utilized autocorrelation times provided by the built-in functionality of the \texttt{emcee python} package \citep{2013PASP..125..306F}. Over 15,000 steps, the maximum autocorrelation time for our five free parameters was 172 steps. So, our model was run for roughly 87 autocorrelation times and was sufficiently converged (runs $\gtrsim$ 50 autocorrelation times have usually converged adequately; see \texttt{emcee} documentation\footnote{\url{https://emcee.readthedocs.io/en/stable/tutorials/autocorr/}}).

The first 344 steps in the sample chains were discarded as a burn-in period---time set aside to allow the walkers to approach a solution prior to sampling their distribution. This burn-in value was selected to be twice the maximum autocorrelation time (see \texttt{emcee} documentation\footnote{\url{https://emcee.readthedocs.io/en/stable/tutorials/monitor/}}). Examination of the trace of the walker paths shows that this is sufficient time to approach a solution.

We compute the reduced $\chi^2$ metric, with values near 1 indicating good fits, for the fit to be 0.27 (slightly over-fit, which is sufficient for this analysis) using the Kraft, Burrows, and Nousek method of calculating a Poisson confidence interval \citep{1991ApJ...374..344K} for the rolling histogram error on the real object distribution. The \texttt{poisson\_conf\_interval} function from the \texttt{astropy.stats python} package \citep{astropy} was used to calculate these confidence values with a background noise of zero and a confidence level of 0.68 ($1\sigma$ assuming Gaussian statistics).

\begin{figure*}
  \centering
  \includegraphics[width=\textwidth]{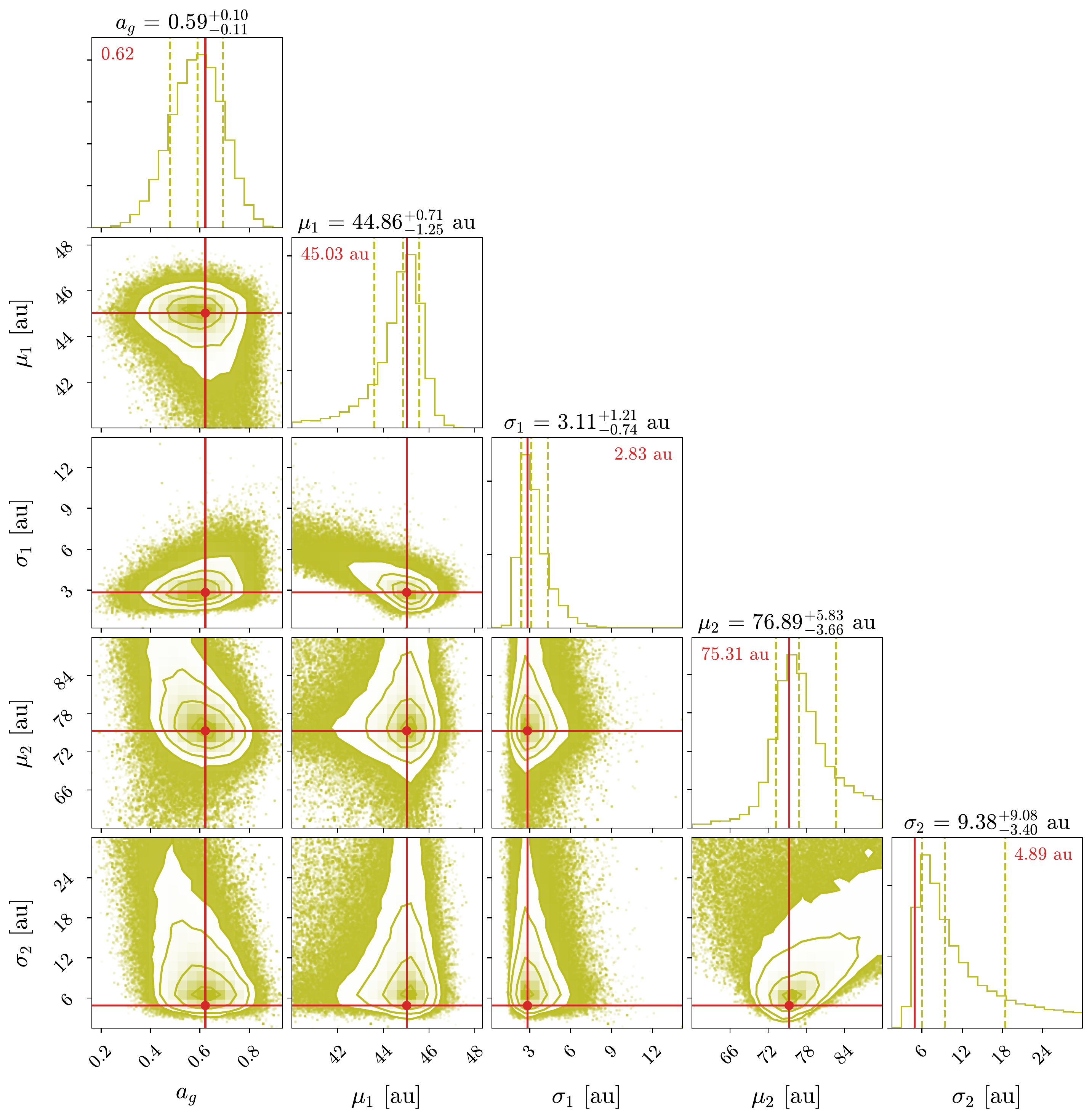}
  \caption{Corner plot of the MCMC sample parameter distributions. One-dimensional histograms for the sample chains of each free parameter are located at the top of each column. Two-dimensional histograms of each possible parameter combination are also shown. Values shown above each column are the model fit reported in Table \ref{tab:mcmc} and correspond to the median of the distribution with uncertainties at the upper and lower 1$\sigma$ quantiles (dashed vertical lines). Solid red lines and their corresponding values within the one-dimensional histogram subplots are the maximum likelihood $\mathcal{L}$ values in Table \ref{tab:mcmc} and result from a single correlated link.} 
  \label{fig:corner}
\end{figure*}

In Figure \ref{fig:corner} we provide a corner plot of the sample distributions of each of the free parameters used in our MCMC run. The values shown correspond to those reported in Table \ref{tab:mcmc}. The solid red lines indicate the parameter values from the correlated link with the highest likelihood value and all of these values are near the peaks of their corresponding distribution; however, the maximum likelihood value for $\sigma_2$ lies just outside of the lower quantile of its distribution. To confirm that this would not have an adverse effect on our results, we tested other values near these peaks using the observational model outlined in Sections \ref{subsec:params} and \ref{subsec:stats}. We find that nearby values (within the given uncertainties) also result in fits that are $\sim$7 orders of magnitude better than the single component continuous distributions we tested (see Table \ref{tab:distributions}).

\section{Dynamical Simulation Animations} \label{app:animations}

\begin{figure*}
\begin{interactive}{animation}{movie218.mp4}
\includegraphics[width=\textwidth]{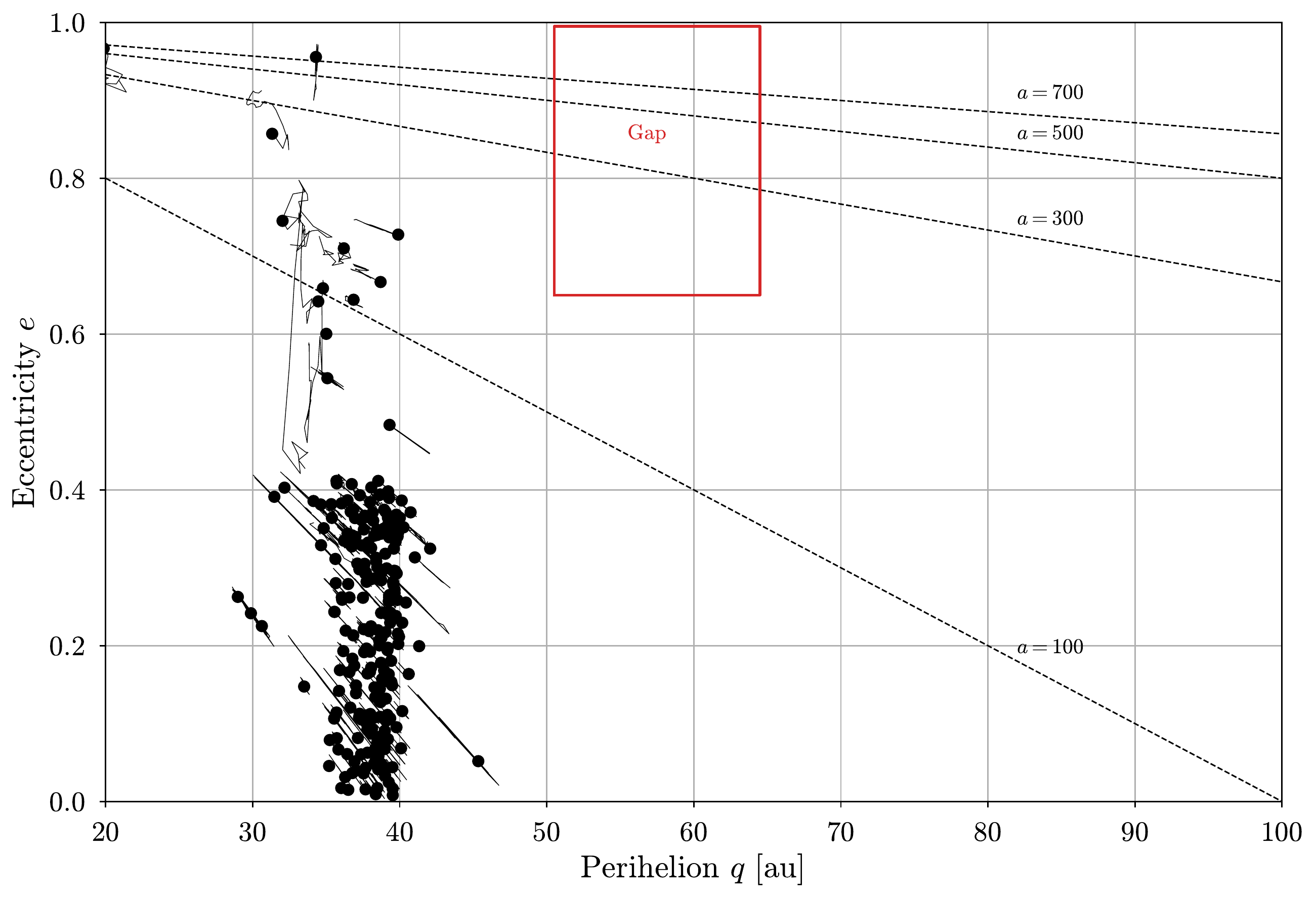}
\end{interactive}
\caption{Animation over 4.5 Gyr of integration of test particles (black points) in eccentricity and perihelion with a Kuiper-Belt-like initial distribution in the presence of the Sun, the known giant planets, and no Planet X. The animation time step is 10 Myr and the trails connected to the test particles track their motion over the previous 250 Myr in the simulation. As the simulation progresses, particles are scattered to high eccentricities ($e>0.6$) through interactions with Neptune. Many particles are captured into mean-motion resonance with Neptune, as expected, and oscillate along diagonal lines of constant semimajor axis. Note that no particles migrate to the ETNO, gap, or IOC regions. The retained Kuiper-Belt-like initial distribution, particles scattered to high $e$, particles in mean-motion resonance, and the lack of particles near the gap are all shown in the single-frame version of the figure. The average particle densities over this simulation are shown as a heatmap in Figure \ref{fig:hm218}.\\ \\(An animation of this figure is available.)}
\label{ani:218}
\end{figure*}

\begin{figure*}
\begin{interactive}{animation}{movie220.mp4}
\includegraphics[width=\textwidth]{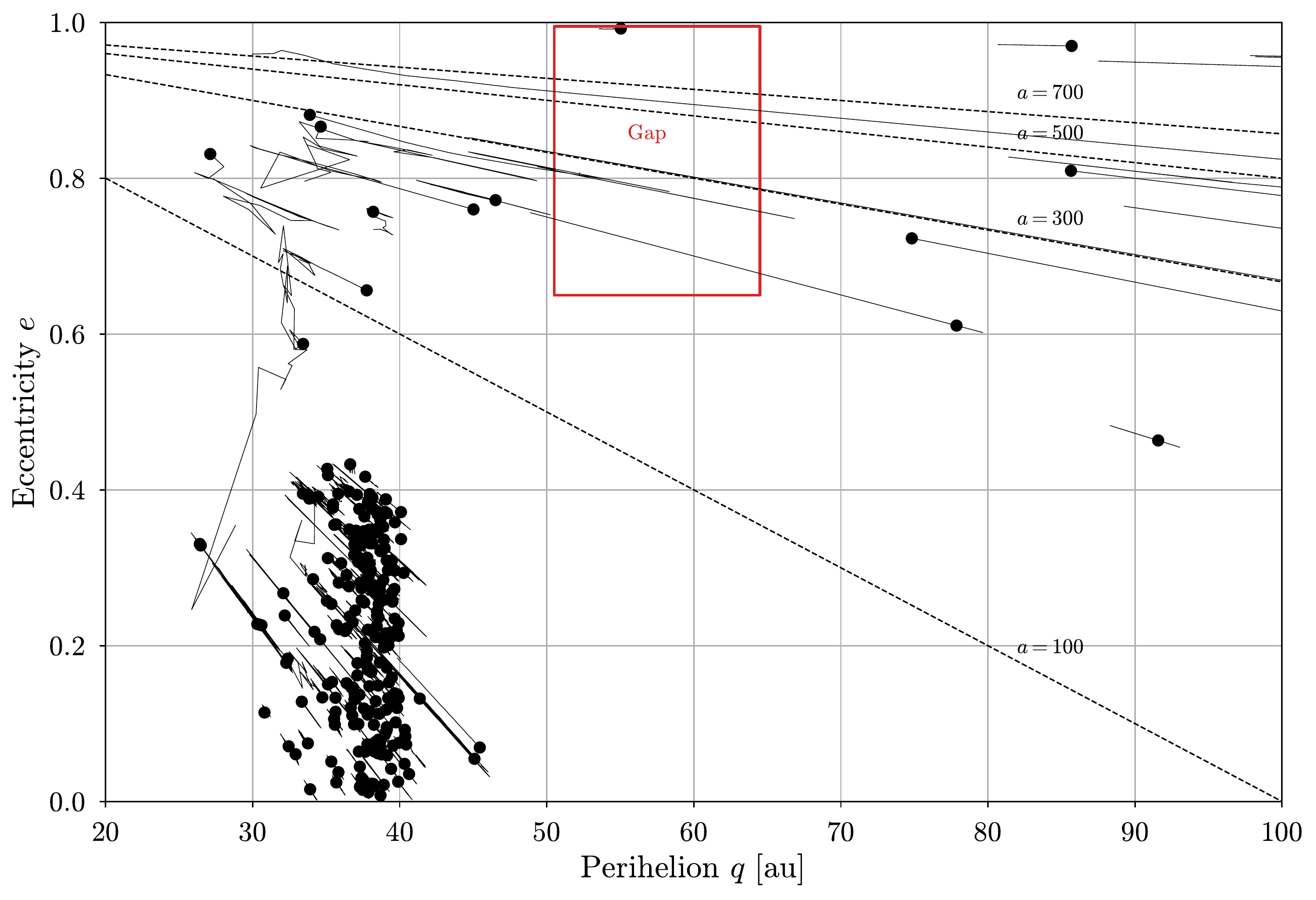}
\end{interactive}
\caption{Animation over 4.5 Gyr of integration of test particles (black points) in eccentricity and perihelion. As Figure \ref{ani:218}, with the addition of Planet X B19 from Table \ref{tab:pxparams} \citep{2019PhR...805....1B}. The animation time step is 10 Myr and the trails connected to the test particles track their motion over the previous 250 Myr in the simulation. Throughout the simulation, particles are scattered to high eccentricity by Neptune (as in Figure \ref{ani:218}), however, many of these particles are now captured into resonance with Planet X and migrate through the ETNO region. Once these particles approach the gap, they migrate more quickly and rapidly transition through the gap into the IOC region and continue beyond $q=100$ au. Particles with eccentricities $e\gtrsim0.95$ migrate through the gap much more slowly. Additionally, many particles return to the IOC region from beyond $q=100$ au. Some of these particles cross back over the perihelion gap into the ETNO region, but others reach the peak of their oscillation cycle within the IOC region. The retained Kuiper-Belt-like initial distribution, particles scattered to high $e$, quickly migrating particles in resonance with Planet X, slowly migrating high $e$ particles in the gap region, and ``returning IOCs" are all shown in the single-frame version of this figure. The average particle densities over the simulation are shown as a heatmap in Figure \ref{fig:hm220}.\\ \\(An animation of this figure is available.)}
\label{ani:220}
\end{figure*}

\begin{figure*}
\begin{interactive}{animation}{aemovie221.mp4}
\includegraphics[width=\textwidth]{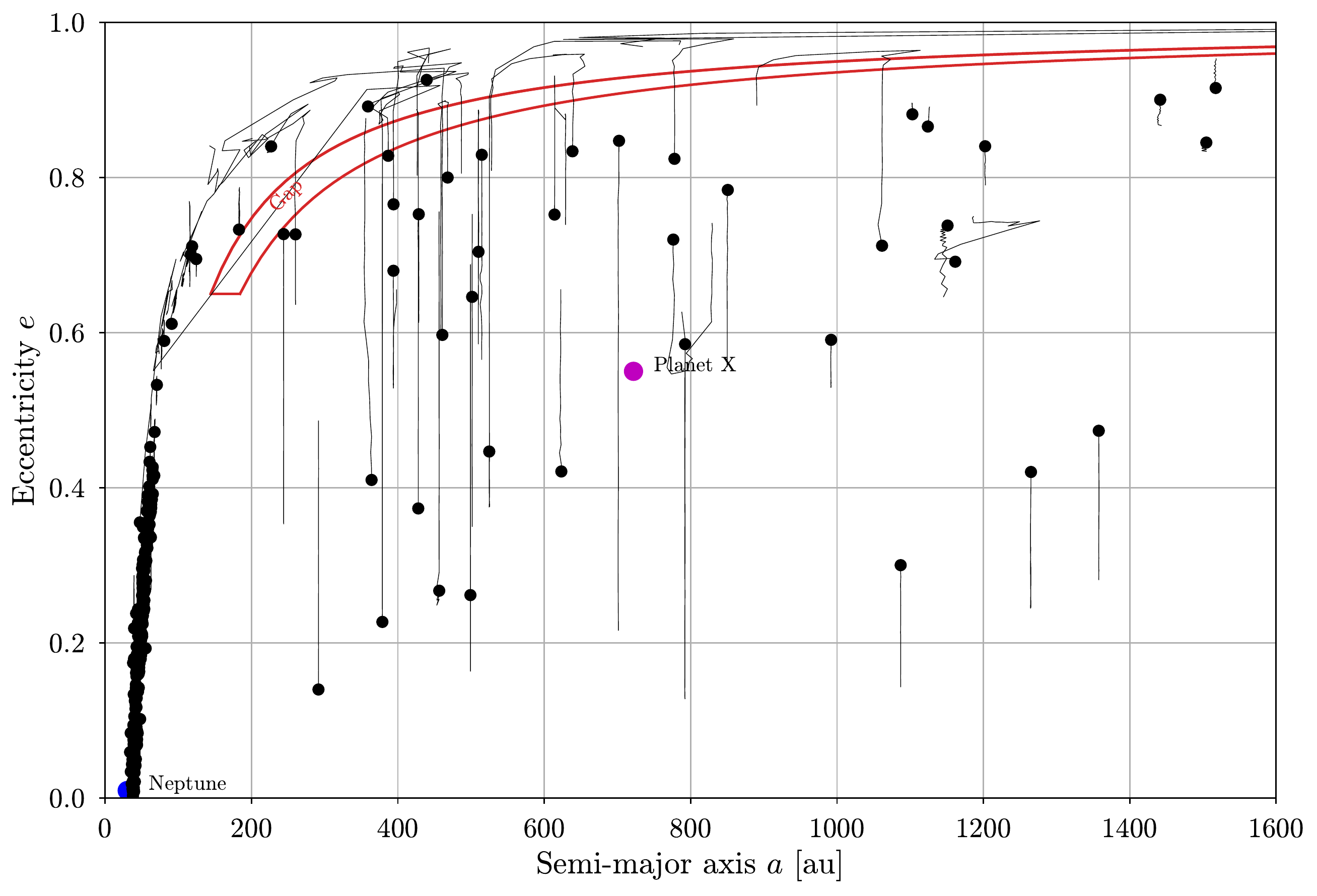}
\end{interactive}
\caption{Animation over 4.5 Gyr of integration of test particles (black points) in eccentricity and semimajor axis with a Kuiper-Belt-like initial distribution in the presence of the Sun, the known giant planets, and Planet X T20 from Table \ref{tab:pxparams} \citep{2020tnss.book...79T}. The animation time step is 10 Myr  and the trails connected to the test particles track their motion over the previous 250 Myr in the simulation. Over the course of the simulation, particles are scattered to high eccentricity by Neptune which also increases their semimajor axes. Many of these particles are captured into secular resonances with Planet X, causing them to oscillate in eccentricity along lines of roughly constant semimajor axis. Occasionally, particles undergo large shifts in semimajor axis during this oscillation period, particularly upon re-entry of the ETNO region where they are again subject to the gravitational influence of Neptune. Additionally, ``returning IOCs" are visible in the simulation, with some crossing the gap toward high $e$ before reaching their peak eccentricity, and others reaching their peak $e$ within the IOC region. The retained Kuiper-Belt-like initial distribution, particles scattered to high $e$, particles in secular resonance with Planet X, particles undergoing ``resonance hopping," and ``returning IOCs" are all shown in the single-frame version of this figure. The average particle densities over this simulation are shown as a heatmap in Figure \ref{fig:aehm221}.\\ \\(An animation of this figure is available.)}
\label{ani:221}
\end{figure*}

Figure \ref{ani:218} is an animation of the particles mapped in Figure \ref{fig:hm218} and highlights how Neptune is unable to form ETNOs, IOCs, or ``gap objects" from a simple Kuiper-Belt-like initial distribution. Figure \ref{ani:220} shows a similar animation, but for the particles mapped in Figure \ref{fig:hm220}. This simulation includes Planet X B19 from Table \ref{tab:pxparams} \citep{2019PhR...805....1B} and shows how objects transition from the ETNO region to the IOC region and beyond via the perihelion gap. ``Returning IOCs" are also apparent in this animation. Figure \ref{ani:221} is an animation of the particles mapped in Figure \ref{fig:aehm221} and contains Planet X T20 from Table \ref{tab:pxparams} \citep{2020tnss.book...79T}. This animation highlights secular resonances of objects with Planet X and their movement along nearly constant lines of semimajor axis.

\end{document}